\documentclass[a4paper]{issibook}
\usepackage[dvips]{graphicx}
\usepackage{amsmath}
\usepackage{natbib}

\newcommand{\abs}[1]{\ensuremath{\lvert #1 \rvert}}
\newcommand{\lvecc}[2]{\ensuremath{( #1 , #2 )}}
\newcommand{\rvecc}[2]{\ensuremath{
              \left( \begin{array}{c} \!\!#1\!\! \\
                                      \!\!#2\!\! \end{array} \right) }}

\renewcommand\thesection{\arabic{section}}

\begin{document}
  
\begin{center}{\Large\bfseries
Stereoscopy basics for the STEREO mission
}\vspace{1ex}\end{center}

\chapauthor{Bernd Inhester\\
            Max-Planck-Institute for Solar System Research\\
            37191 Katlenburg-Lindau, Germany}
\ISSInote{draft version,\qquad 8$^\mathrm{nd}$ August 2006,
                        \qquad to appear as an ISSI publication}

\hspace*{\fill}
\parbox{13cm}{
We discuss some basic principles of stereoscopy and their relevance to
the reconstruction of coronal loops. The aim of the paper is to make
the solar physicist familiar with basic stereoscopy principles and to
give hints how they may apply to the analysis of data from the
forthcoming STEREO mission. We disucss the geometry of the solar
coronal stereo problem, give the basic principles of a tie-point
reconstruction algorithm and consider ambiguities and resolution
errors. Finally we mention extensions to plain stereoscopy such as a
third view, a tomography-like approach and how magnetic field
information can be used to improve the reconstruction.}
\hspace*{\fill}

\section{Introduction}\label{ch:intro}

One of the first applications of stereoscopy in space and
astrophysics was the determination of the distance of near-by stars by
measuring the parallax angle. Even though stereoscopy is now applied
to much more complex objects, the essential reconstruction principles
still remain the same.
Once an object is detected and identified in two images from different
vantage points, the reconstruction is a purely linear geometrical task.

However, the most challenging problem associated with stereoscopy
precedes the geometrical reconstruction step: the identification and
matching of the objects to be reconstructed in the stereo images. But
even before we start to talk about identification, matching and
reconstruction, we have to be concerned about which are the objects
that we hope to find in the images.

This text is strongly devoted to solar physics and especially meant as
background for STEREO mission scientists. On the other hand, our
intention here is to draw parallels to and learn from other fields of
physics and engineering where stereoscopy has been applied for a long
time.

The classical stereoscopy problem is the reconstruction of surfaces
from a pair of images. For complicated surfaces such as, e.g., human
faces this is a formidable task and not possible without well
calibrated images and highly specialized information beyond the image
data alone, e.g., the location of light sources and the reflectivity
of the human skin.

A more elementary task is the reconstruction of simple geometrical
objects composed of piecewise planes connected by straight edges. The
reconstruction in this case can already be achieved if the only edges
can be identified in the images. They usually show up as sharp
boundaries in brightness, colour or texture -- the absolute values of
the brightness, colour or texture are not needed. Subsequently, the
edges are reconstructed first and hopefully, suitably connected
edges lie in planes which then bound the desired surfaces.

\begin{figure}
  \hspace*{\fill}
\includegraphics[width=13cm]{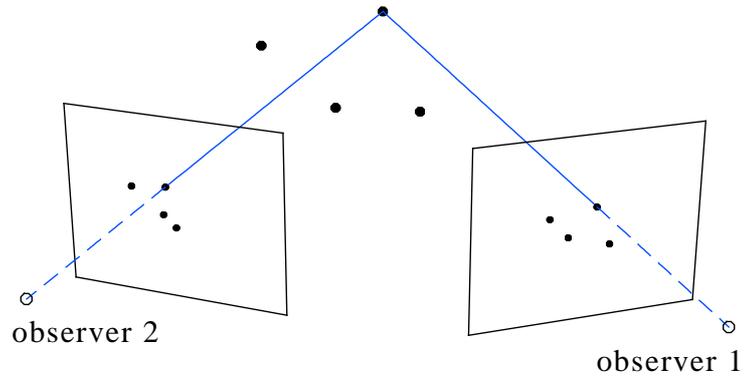}
  \hspace*{\fill}\\
  \hspace*{\fill}
\includegraphics[width=13cm]{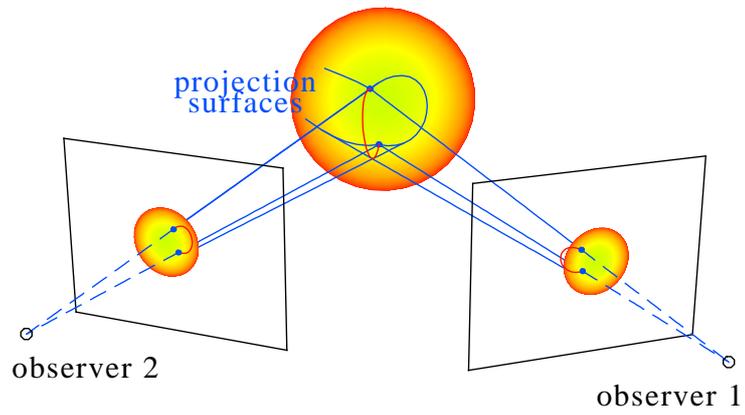}
  \hspace*{\fill}\\
  \hspace*{\fill}
\includegraphics[width=13cm]{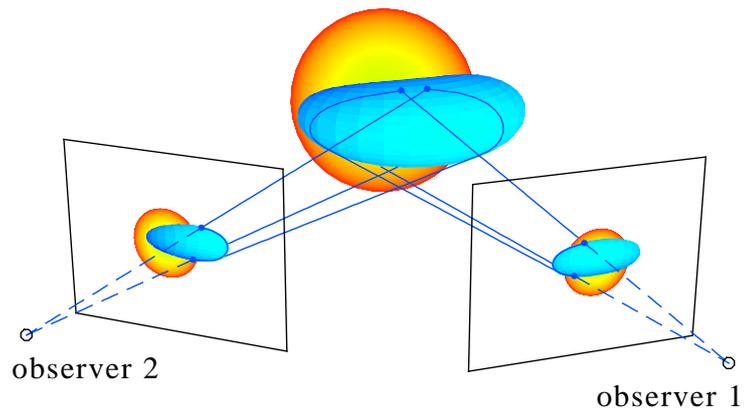}
  \hspace*{\fill}
\caption{Backprojection to reconstruct point-like, curve-like and
  surface-like objects to demonstrate the different conditions of
  solvability.
  \label{fg:dimensions}}
\end{figure}

There is a good reason why the edges are the natural objects to be
reconstructed first.
If have identified an object, the reconstruction task, i.e., the
calculation of its depth, depends on the extent or dimension of the object:
\begin{list}
  {$\bullet$}
  {\setlength\topsep{0pt}%
   \setlength\partopsep{0pt}%
   \setlength\parsep{0pt}%
   \setlength\itemsep{0pt}%
   \setlength\leftmargin{2em}%
   \setlength\rightmargin{2em}}
\item For a point-like object, we extract 2$\times$2 image coordinates
  from the image pair from which we determine by a linear relationship
  the 3 space coordinates of the object -- a clearly overdetermined
  mathematical problem (Fig.~\ref{fg:dimensions}a).
\item For a curve-like object, we extract two two-dimensional curves
  from the image pair, each of which could be considered the
  ``head-on'' image of a projection surface extending in the
  respective view direction. The intersection of the two projection
  surfaces yields (in the ideal case) a unique three-dimensional
  curve, the desired result of our reconstruction
  (Fig.~\ref{fg:dimensions}b).
\item For a surface-like objects, we can its extract visible edges from
  the image pair, but as we see in Fig.~\ref{fg:dimensions}, the
  respective projection surfaces refer to different locations on the
  surface to be reconstructed. Our problem is obviously
  underdetermined and without further information, e.g., the
  assumption of a surface curvature or the knowledge about the
  location of a light source, the surface reflectivity together with
  precise intensity measurements, a reconstruction is impossible
  (Fig.~\ref{fg:dimensions}c).
\end{list}

Hence there is a good reason why in classical stereoscopy the edges,
which are curve-like objects, are reconstructed first.
On the solar surface similar objects which lend themselves to being
reconstructed are striated plasma structures extending along the
coronal magnetic field.
Already the first applications of stereoscopy in solar physics by
\citet{BertonSakurai:1985} and \citet{KouchmyMolodensky:1992} considered
active region loops and plasma striations in white-light coronagraph
observations.

The solar corona is dominated both dynamically and statically by its
magnetic field. The visible striations above the solar surface are
essentially shaped by the efficient transport properties along the
field lines and are therefore smoothly curved.
When talking about three-dimensional curves as objects of
reconstruction, we will have these plasma structures in mind.
Below we will explicitly discuss loops, which are plasma striations on
presumably closed magnetic field lines, and plumes, i.e., elongated density
enhancements on open field lines, visible over the
polar coronal holes preferentially at times of the solar activity minimum.

The loops clearly show up in EUV images of the solar corona since the
about 1 million K hot coronal plasma confined on thin flux tubes
emits much stronger in EUV wavelengths then the cool solar surface of
only 5600 K. The high contrast is probably due to both drastic temperature
and density variations across coronal field lines. Note that the resulting
variation in pressure is still small compared to the shape-forming forces
of the magnetic field.
The huge temperature difference between the corona and the lower
atmosphere layers largely eliminates backround radiation from these
lower layers so that the large intrinsic contrast of coronal loops
is preserved even though the plasma density in the corona is by
orders of magnitude smaller than the density at the surface.

Plumes have less contrast but are seen in optical and EUV wavelengths,
probably, because their enhanced emission is due to an enhanced
density on some polar field lines an not so much due to a drastic
temperature change. Best contrast is obtained where the plumes extend
beyond the visible solar disk and can be observed against the dark
background sky.

We will start with introducing the natural geometry attached to
stereoscopic observations. The epipolar coordinate system thereby
introduced is already a first step towards the important task to
associate features from both images with each other. This task, as
mentioned above, is formidable and is outlined in the third section.
Much of the research in various fields of stereoscopy is devoted
to this step and especially for solar observations there is still much
work ahead on this special problem.
The geometrical reconstruction task itself is elaborated in some depth
in the tie-point method chapter. Even the geometrical reconstruction
can fool the unaware user and the topics of the following sections
deal with the precision of the reconstruction and the uniqueness of
whatever we hope to obtain as final result.
Additional observations, e.g., a ``third eye'', magnetic field
observations, and alternative methods and constraints to the
geometrical reconstruction such as the tomography approach are
discussed at the end.

All these latter methods together with our basic knowledge of physics
should be seriously considered as additional potential constraints
when analysing images whenever possible.
We should keep in mind that the step from images, i.e., 2D projections
to the 3D real world is often underestimated. As humans we percieve
our 3D environment largely through the images produced on the focal
plane of our eyes. We have calibrated our sense of depth in early
childhood and we have for years classified the objects which surround
us in every day life. Whenever we see a projection of one of these
objects we immediately associate this limited projection with the
3D-idea of this object we have stored in our brain. Most often we are
not aware any more of this process and we therefore tend to consider
2D projections and their 3D shape as equivalent. But if we want to
teach a computer to make a 3D reconstruction, we have to give this
additional information to him explicitly. Even worse, in space we
still largely lack this important background of information and we so
find ourselves in this area still ``in the childhood of perception''.

\section{
Epipolar geometry and Rectification}\label{ch:epipgeom}

The first step is to find a suitable coordinate system so that the
reconstruction can be reduced from a three-dimensional to a set of
two-dimensional planar problems.
A similar strategy is known from tomography where the reconstruction
can be obtained first for a dense set of planes normal to the rotation
axis of the object (or the camera). The final three-dimensional model is
then interpolated in between these planes. A necessary condition for
the validity of this procedure is that the rotation axis lies normal
to the plane of view directions and that affine geometry for the
camera projection rays can be assumed.

\begin{figure}
\hspace*{\fill}
\includegraphics[width=12cm]{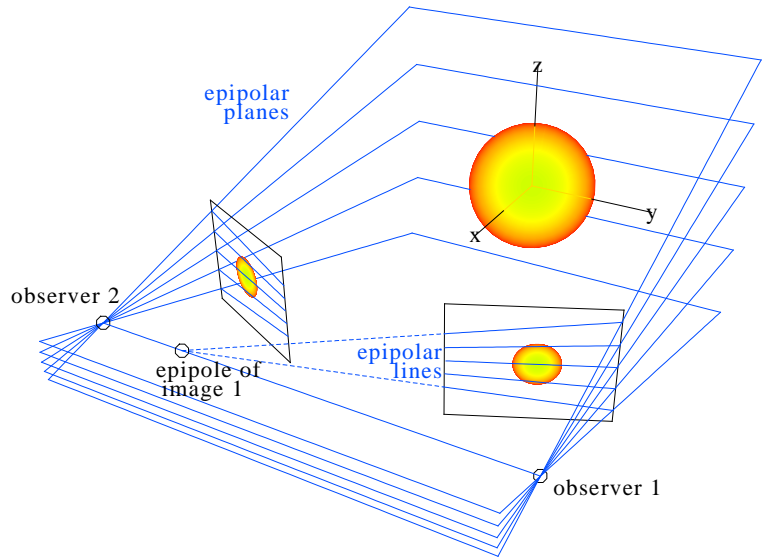}
\hspace*{\fill}\\
\caption{Orientation of epipolar planes in space and the respective
  epipolar lines in the images for two observers (e.g., space craft)
  looking at the Sun. The observers telescope screens are derived from
  a projective geometry camera model.
  \label{fg:epiplanes}}
\end{figure}

For stereoscopy the geometrical conditions for this segmentation of
space are much more relaxed because only two view points and two view
directions are involved.
The line connecting the view points is called the {\bf stereo base line},
which subtends the {\bf stereo base angle} between the two main view
directions (more precisely, the optical axes of the respective
telescope. Note that the two optical axes do not need to intersect).
The two observer positions and any object point to
be reconstructed exactly define a plane. For many object points there
are many planes but all have in common that they contain the two
observer positions. Any of this set of planes which contains the two
observer positions is called an {\bf epipolar plane} and these planes form a
natural geometrical basis for our reconstruction coordinate system
(see Fig.~\ref{fg:epiplanes}).

By the above definition, epipolar planes project on both observer's
images as lines: since any observer himself is on any epipolar plane,
he sees them ``head on''. These lines are called accordingly the
{\bf epipolar lines} and they generate a natural coordinate system on the
image planes.

Depending on his field of view, one observer, say observer 1, may see
observer 2. Since observer 2 is defined to lie on all epipolar planes,
all epipolar lines in image 1 must converge in the projection of
observer 2, called the {\bf epipole} of the respective projection.

The epipolar lines in each image can easily be determined from the
observers positions and the direction of their optical axes. Note that the
epipolar lines in one image always depend on the position of both
observers, hence any change in position of observer 2 requires a
redetermination of epipolar lines also in image 1. The epipolar lines
may be labelled by the angle they make with respect to a reference
epipolar plane or by the coordinate of the plane's intersection with a
suitable axis of a completely independent coordinate system, e.g., the
$z$-axis in Fig.~\ref{fg:epiplanes}. In the STEREO context, a
suitable axis is the rotation axis of the Sun. Since the space craft
are more or less close to the ecliptic, the Sun's rotation axis should
intersect all relevant epipolar planes. This way, an epipolar plane is
uniquely defined by the two observer's positions and the intersection
point.

Conversely, most space points lie on a unique epipolar plane.
The only exception are points on the stereo base line which
connects the two observers.
In fact there is no depth reconstruction possible for points on this
line. If we rule them out, any point identified in one image on a
certain epipolar line must occur on the same epipolar line in the
other image.

This straight forward geometrical consequence is known as {\bf epipolar
constraint}. It enormously helps the reconstruction since, in principle,
we can directly overplot the epipolar lines and their labels onto each
image and thus read the epipolar plane of an object from the image.
Hence, we are only left with calculating the object's two-dimensional
coordinates on the known epipolar plane. This can be achieved
from measuring the positions of the object's projection along the epipolar line
in either image. The depth of an object in particular is proportional
to the difference of these coordinate values along the epipolar line
of the two projections, the so-called {\bf disparity} (for a quantitative
definition see section \ref{ch:tiepoint}).

From the geometrical construction it is clear that epipolar lines
usually are not parallel. In the case of the STEREO spacecrafts at
about 1~AU the visible epipolar lines typically make angles of
$\mathcal{O}(\mathrm{atan}~1 R_{\odot})/1 \mathrm{AU}$ $\simeq$ 0.01
in an EUVI image concentrated on the solar disk. For the HI images,
on the other hand, the epipoles come into the field of view and
epipolar lines in the image may make angles of $\mathcal{O}(1)$.

\begin{figure}
\hspace*{\fill}
\includegraphics[bb=0 60 580 430 ,clip, width=12cm]{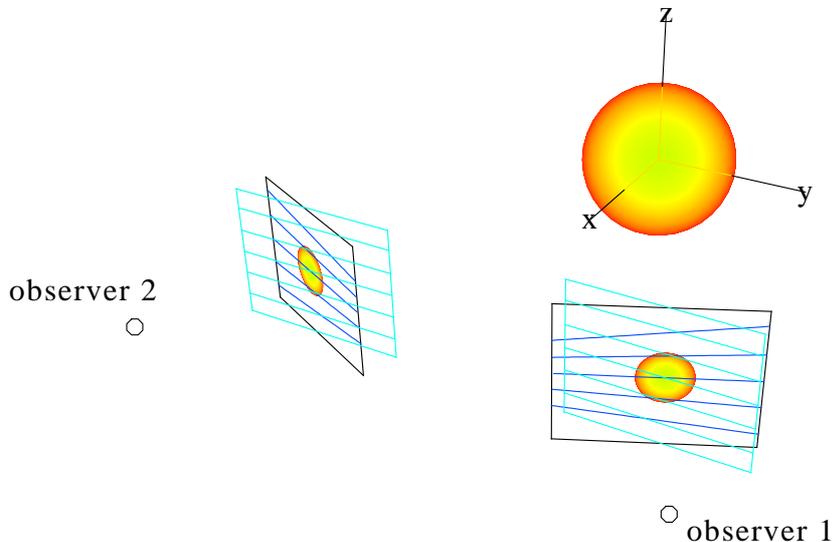}
\hspace*{\fill}\\
\caption{Virtual change of orientation of the observer's main view
  direction equivalent to rectification. The rectified configuration
  is indicated by the rotated screens drawn in light blue.
\label{fg:rectify}}
\end{figure}

For ease of reconstruction it would be desirable to have the epipolar
lines mapped into horizontal lines. This corresponds to the case when
both spacecrafts would have their optical axes directed parallel to
each other (see Fig.\ref{fg:rectify}). The transformation of a
general image pair to an pair with horizontal epipolar lines is called
{\bf rectification}. Usually, this step includes also the elimination
of distortion effects from the image. Note that in presence of a
distorted image mapping, epipolar lines may not be straight in the
original image.
In rectified images, a space point maps to the same
vertical image coordinate in both images. Hence the image
reconstruction can be performed row by row.

Whether the stereo image pair is rectified or not -- the epipolar
constraint is a strong geometrical constraint and has to be obeyed
whatever reconstruction algorithm is employed. Using the epipolar
coordinates explicitely reduces the three-dimensional reconstruction
problem to a set of two-dimensional problems.

\section{
Identification and Matching}\label{ch:matching}

The matching problem consists of correctly identifying the projections
of an object in the two stereo images. The relationship thus established
between the projections is called a {\bf correspondence}.
We will discuss further below in section~\ref{ch:ambiguities} what the
consequences are if the problem is not solved properly before the
reconstruction is made. In general, ghost signals will appear and
one possible approach is to perform the matching imperfectly as
good as possible and select remaining ghosts afterwards in the
reconstructed three-dimensional model. This post-processing selection
must then be based on a-priori known properties of the reconstructed
objects -- we hence incorporate additional information beyond the image
data into our problem. To some extent, this approach will be discussed
below in sections~\ref{ch:ambiguities} and \ref{ch:magnetic}.

On the other hand, it would of course be desirable to add this
knowledge already at the matching stage before the reconstruction is
made. Therefore solutions to the matching problem are often strongly
tailored to the specific objects observed.
EUV and coronagraph images of the solar corona often show a mess of
loops and structures and their identification is a hard task indeed. 
For the STEREO mission, these rules and tools still have to be
developed to a large extent and we can give here only some
general recommendations.

The methods to find correspondences between the stereo images
used in classical stereoscopy are generally classified
in correlation- and feature-based approaches
\citep[e.g.,][]{TruccoVerri:1998}.
In the former method a cross correlation is calculated between the
intensity of the two stereo images along common epipolar
lines.
Maxima of this correlation as function of the relative shift along the
epipolar line are taken as evidence for a local correspondence. The shift
value at which this correlation maximum occurs is collected in
disparity maps from which depth information can be retrieved.
This method has proven successful where depth varies smoothly over the
image and where surface texture and a short stereo base allow firm
correlation maxima to be established.

The plasma striations we are looking for with the STEREO telescopes
stick out of or hover above the solar surface and obviously they do not
well yield the depth continuity provided by a smooth surface.
For this reason it seems that the alternative feature-based method
to establish correspondences is more adequate for our problem.
With this method the correlation is essentially replaced by a search problem:
For a feature on a given epipolar line in one image, we scan the same epipolar
line in the other image to find the corresponding feature there. The
disparity now is the difference in the coordinate along the epipolar line
of corresponding features.

The feature in our case is the intersection of a loop or plume with an
epipolar line. It could be desirable therefore to reduce the image
content to just the loop projection curves. This image segmentation is
a problem of its own and will be dealt with in a separate
contribution. The ideal result of the segmentation step would be a set
of image curves which indicate the centre trace of the projected loops or
plumes in the image (\citet{Strous:2002, LeeJKEtal:2006}, CLAW code
of Jean-Francois)

For feature-based correspondences, however, we still need some guidance which
tells us which correspondence between the intersections along an
epipolar line in the stereo image pair are the most probable. These rules
can loosely be grouped into
1) rules which tell us when to discard certain correspondences and
2) similarity measures which evaluate the probability of a correspondence.
Below we give a list of some of these rules and measures and to
which extent they are useful for coronal EUV and coronagraph images.
This list, especially concerning the similarity measures, is by no means
exhaustive and in future we need an extension of this list to be derived
from our experience with the STEREO data.\\
1)~Rules to discard certain correspondences:
\begin{list}
  {$\bullet$}
  {\setlength\topsep{0pt}%
   \setlength\partopsep{0pt}%
   \setlength\parsep{0pt}%
   \setlength\itemsep{0pt}%
   \setlength\leftmargin{2em}%
   \setlength\rightmargin{2em}}
\item {\bf Uniqueness constraint}.\\
  If we determine a certain feature in image 2 as the most probable
  correspondence for a feature in image 1, we should find the same
  correspondence the other way around: the feature in image 1 should
  turn out as the most probable for the feature in image 2.
  Correspondences which are not unique in this sense have to be
  rejected.
\begin{figure}
\hspace*{\fill}
\includegraphics[bb=30 30 330 260, clip, width=10cm]{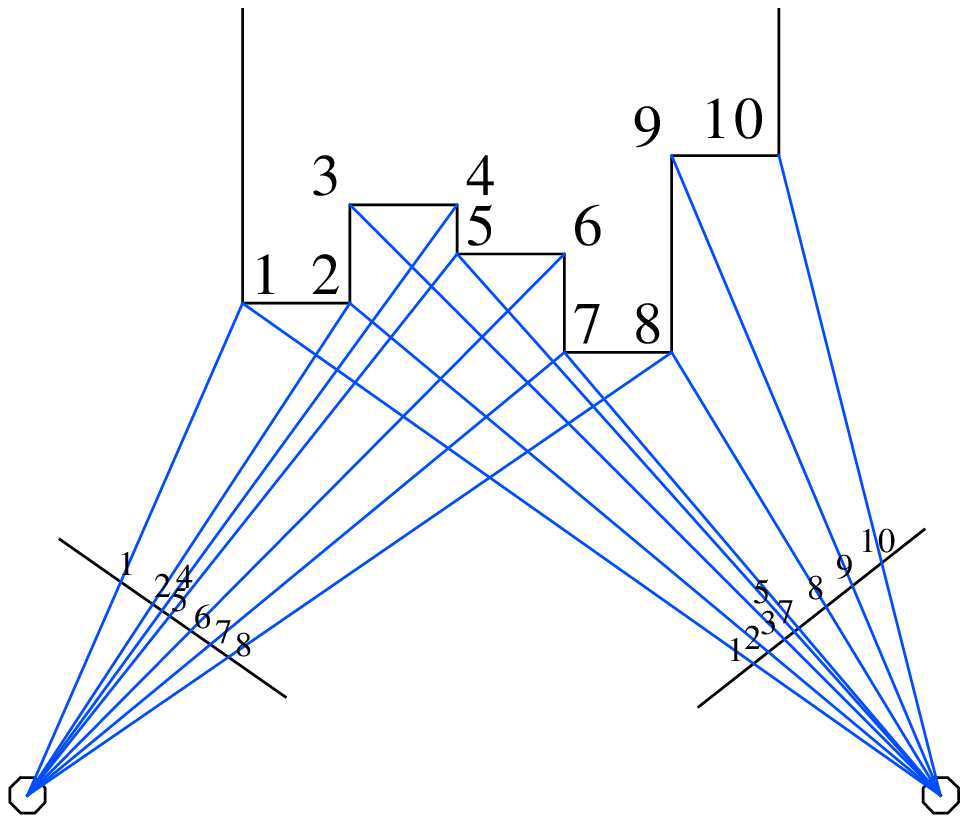}
\hspace*{\fill}\\
\hspace*{\fill}
\includegraphics[bb=30 30 330 260, clip, width=10cm]{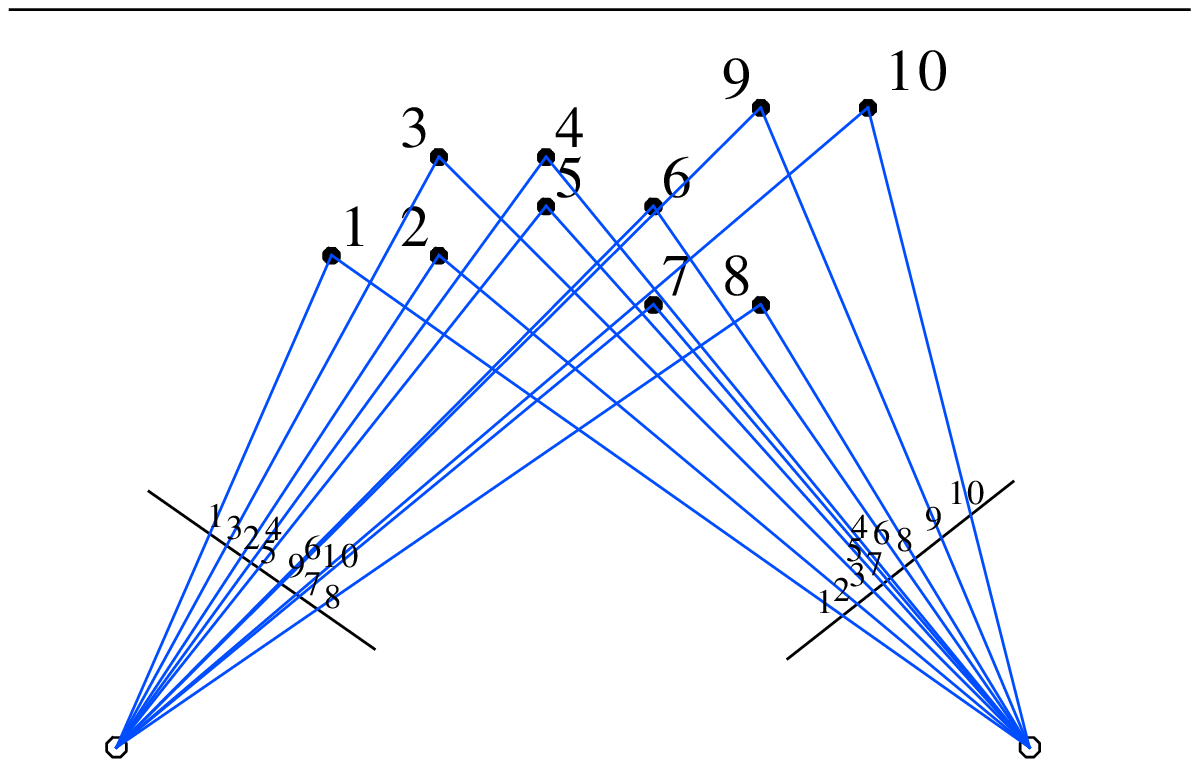}
\hspace*{\fill}
\caption{Illustration of the ordering constraint in classical stereoscopy.
  The top sketch shows an epipolar plane intersected by a piecewise
  planar surface. The edges and their projection on the stereo images
  are numbered to keep track of the viewing perspective.
  The bottom sketch shows the epipolar plane if the edges were replaced by
  isolated loops intersecting the epipolar plane.
\label{fg:ordering}}
\end{figure}
\item {\bf Ordering vs. completeness constraint}.\\
  When looking at a continuous, intransparent surface from different
  view-points, the order at which, e.g., edges are seen in any image
  along an epipolar line is unchanged. It may happen though that an
  edge is not visible at all because it is hidden from one of the view
  points. Fig.~\ref{fg:ordering}a illustrates this ordering
  invariance. Any set of correspondences established between the
  projections of the edges has to obey this rule and it often strongly
  constrains the possible correspondences.
  For the coronal loops and plasma striations, which are optically
  thin and transparent, this ordering constraint cannot be applied.
  Instead, if the stereo base angle is not small, the order at which
  loops are seen along an epipolar line may be permuted (see
  Fig.~\ref{fg:ordering}b).
  To some extent, the ordering constraint is replaced by a
  completeness constraint for solar EUV and coronagraph stereo images.
  In principle, no loop or plasma striation should be lost in the
  images and any object we see in one image should have a
  correspondence in the other image. There are practical limitations,
  however, to this completeness constraint. Some features may lie on
  top of each other, may be lost if they are faint or are hidden
  behind the solar disk if they are located at the edge of the solar
  surface overlap region of the two STEREO spacecrafts.
\item {\bf Disparity limitation}.\\
  For the objects we observe in solar and coronal stereo images we
  have a least a rough idea where they are (or better, where they are
  not) located: not below the solar surface and often we can also give
  an approximate upper limit for the height above the surface. This
  valid height range for an object directly transforms into an allowed
  disparity range. Hence, when searching for a correspondence for a
  feature in one image, the coordinate range along the epipolar axis
  of the corresponding feature in the other image is limited
  accordingly.
\item {\bf Disparity continuity constraint}.\\
  A smoothly varying depth should yield a smoothly varying
  disparity both along epipolar lines and on neighbouring epipolar
  surfaces. If applicable, this constraint could favour certain
  local correspondences if those in the neighbourhood were firmly
  established. As mentioned above, this rule is not useful for the
  loops and plumes we observe in the solar corona. We expect, however,
  that the three-dimensional curves which represent the location of
  these objects vary smoothly. Therefore the continuity constraint
  can be expected to hold at least in the one direction along the curve
  projections.
\end{list}
2)~Similarity measures:
\begin{list}
  {$\bullet$}
  {\setlength\topsep{0pt}%
   \setlength\partopsep{0pt}%
   \setlength\parsep{0pt}%
   \setlength\itemsep{0pt}%
   \setlength\leftmargin{2em}%
   \setlength\rightmargin{2em}}
\begin{figure}[t]
\hspace*{\fill}
\includegraphics[width=12cm]{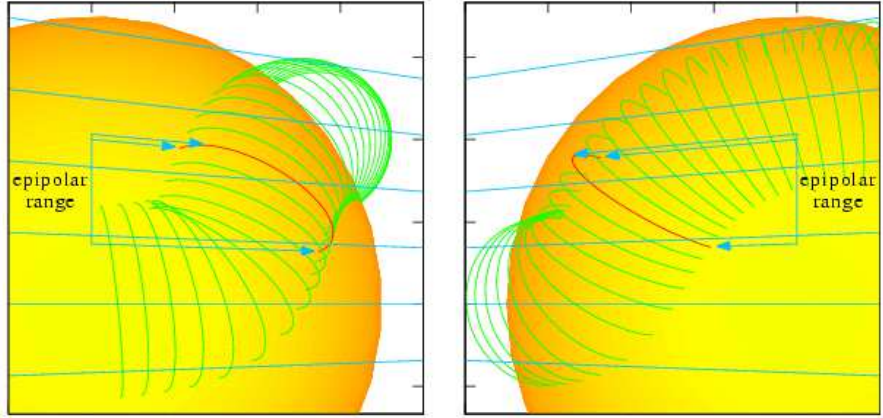}
\hspace*{\fill}
\caption{Identifying loops by their epipolar range. The epipolar
  lines (light blue) intersected by a particular loop must be identical
  in both images.
\label{fg:epirange}}
\end{figure}
\item {\bf Common epipolar range}.\\
  Not only must corresponding features be located on the same epipolar
  line, but if they are part of a projection from, e.g., the same
  loop, the whole loop must cover the same epipolar range in both
  images. Since the visible ends of loops often fade out, it may not
  always be possible to precisely determine its epipolar range. Yet, a
  comparison of the observed smallest and largest epipolar line value
  covered by a loop or plume projection including error ranges could
  help to distinguish loop projections as
  sketched in Fig.~\ref{fg:epirange}. Even more, if a loop projection
  covers a certain epipolar range twice or more in one image, its
  projection in the other image must exhibit the same multiplicity in
  the same epipolar ranges. In properly rectified images this
  comparison of epipolar coverage is particularly straight forward.
\begin{figure}[b]
\hspace*{\fill}
\includegraphics[bb=15 30 410 185, clip, width=10cm]{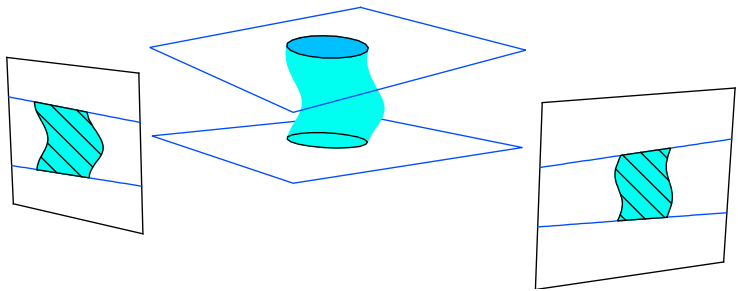}
\hspace*{\fill}\\
\caption{Illustration of the integral intensity preservation.
\label{fg:intpres}}
\end{figure}
\item {\bf Integral brightness invariance}.\\
  If the emission from the coronal plasma striations is isotropic,
  the integrated brightness of a projected loop section (with background
  subtracted) between two epipolar lines should correspond to the
  total emissivity of the three-dimensional loop section between the
  respective epipolar planes and hence should be the same in both
  images (see the related Helgason-Ludwig consistency condition in
  tomography, e.g., \citet[p.\,36]{Natterer:1986}). This rule is immediately
  obvious if we calculate the image intensity by appropriate
  line-of-sight integrals through the emitting volume as displayed in
  Fig.~\ref{fg:intpres}. The rule holds independent of the inclination
  the loops with respect to the epipolar planes. But it only holds if epipolar
  planes are chosen as loop section boundaries: any other boundary
  plane is viewed differently from the two observers.
  The integrated brightness of loop intersections along an epipolar
  line evaluated in both stereo images can therefore effectively help to
  find the most probable correspondence between the intersections in the
  two images.
  A prerequisite for an application of this rule are well calibrated
  images and instruments with similar signal to noise ratio. The
  STEREO mission should provide them. A crucial point may be the
  continuity of the background radiation which has to be subtracted
  from the integrated loop section intensity.
  The similarity measure should therefore allow for small deviations
  of the integrated brightnesses of the order of the background intensity.
\end{list}

More rules can be derived from combinations of different items of the above
list: E.g., from the disparity continuity along a loop and the criteria
which lead to the disparity limitation we can conclude: If the branch
(see section~\ref{ch:ambiguities} for a precise definition) of a
reconstructed loop includes a part which reaches below the solar surface,
we can identify the whole branch as a ghost and reject it.

The above list needs to be extended. We are convinced that the reconstructed
loop's consistency with magnetic field models will also lead to helpful
similarity measures between loop projections. Since this approach requires
the incorporation of additional observations besides the image data,
we discuss details of this approach in a separate section \ref{ch:magnetic}
below.

The permutation of objects along the epipolar line if viewed from
different directions (Fig.~\ref{fg:ordering}b) obviously increases with
increasing stereo base angle -- one reason why the matching is easier
for small stereo base angles and why small base angles are often considered
particularly favourable for stereoscopy.
For the STEREO mission is also important to keep in mind that with
increasing stereo base angle the overlapping part of the solar surface
is reduced. We will see below, however, that the reconstruction error
is much larger for small base angles. Therefore the question for the
optimum base angle is difficult to answer. It depends on the one hand
on the complexity of the objects which make matching difficult and on
the other hand on the desired precision of reconstruction.
For the STEREO mission the whole range of base angles beyond 10
degrees will be scanned with time and we will have the chance to test
our schemes for practically all angles.

\section{
Tie-point reconstruction}\label{ch:tiepoint}

As we learned above, we can reduce the reconstruction problem to a set
of two-dimensional problems. We segment each image densely into a
(large) number of epipolar lines and compare the positions of a loop's
intersections along the respective epipolar line in each image. In this
section we show how we can derive from these positions the
intersection of the loop on the respective epipolar planes.

\begin{figure}[t]
\hspace*{\fill}
\includegraphics[width=8cm]{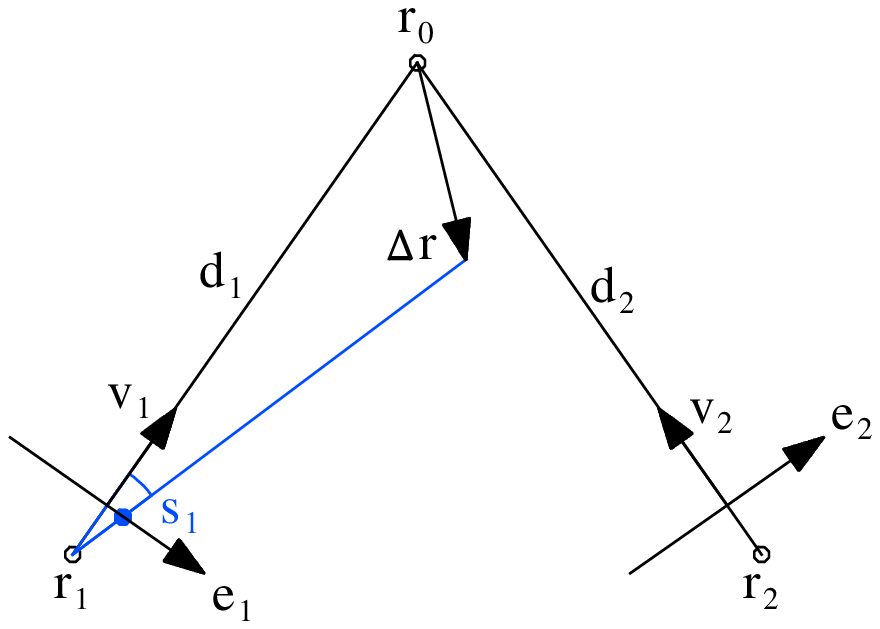}
\hspace*{\fill}\\
\hspace*{\fill}
\includegraphics[width=8cm]{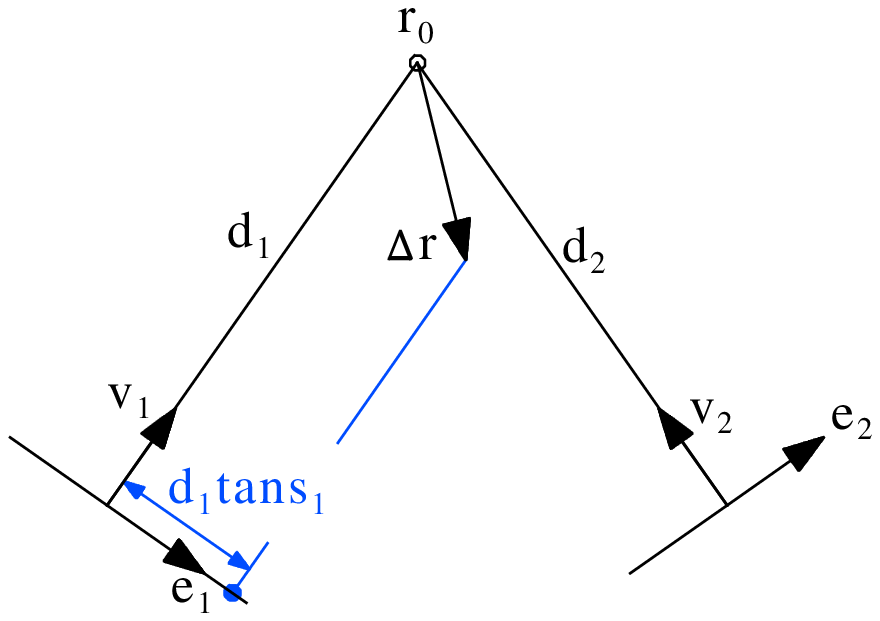}
\hspace*{\fill}
\caption{Reconstruction of a point $\Delta\vec{r}$ with
 projective (top) and affine (bottom) geometry. 
\label{fg:recon}}
\end{figure}

Let on a given epipolar plane the observer's positions be $\vec{r}_1$
and $\vec{r}_2$. For each epipolar plane we in addition specify a
reference point $\vec{r}_0$ as the origin of the two-dimensional
coordinate system on this plane.
For convenience, we could take the intersection of the solar rotation
axis with the epipolar plane the intercept of which could at the same
time serve as a continuous label for the epipolar plane.
For each observer we then introduce orthogonal coordinate axes
$\vec{v}_i$ and $\vec{e}_i$ on the epipolar plane as shown in
Fig.~\ref{fg:recon}: $\vec{v}_i$ is the unit vector from the observer
to $\vec{r}_0$ and $\vec{e}_i$ is $\vec{v}_i$ rotated clockwise
by 90 degrees.
Note that $\vec{v}_i$ do not need to agree with the optical axis of
the telescope nor must $\vec{e}_i$ have the direction of the epipolar line.

\begin{figure}
\hspace*{\fill}
\includegraphics[width=10cm]{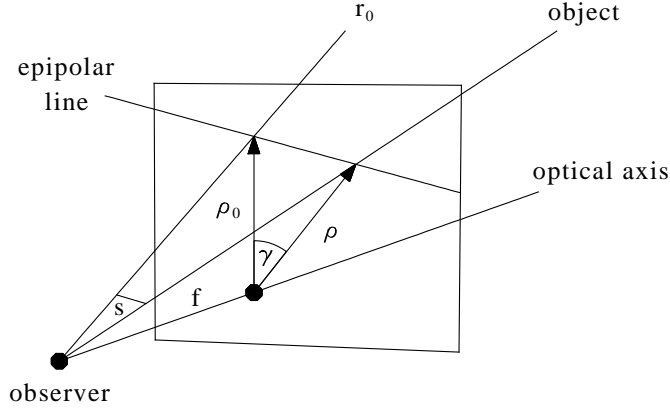}
\hspace*{\fill}\\
\caption{Derivation of the angle $s$ from the image coordinates
  for a projective geometry camera model.
\label{fg:gets}}
\end{figure}

We will take as rectified image coordinate along the respective
epipolar line in image $i$ the angle $s_i$ between the direction
to the object and $\vec{v}_i$.
For convenience, we assume that the mapping of the observing
telescopes can be described by a simple projective geometry camera model.
In this case, an object at an angular distance $\sigma$ from the optical
axis is mapped in the image to a distance $\rho$ = $f \tan\sigma$ where
$f$ is the camera's focal length.
In this case, the angle $s_i$ can be derived from the image distances
$\rho_0$ and $\rho$ of the reference and object point's projection
from the image centre, respectively, and their azimuth difference $\gamma$
(see Fig.~\ref{fg:gets}) by
\begin{equation}
  \cos s_i =\frac{f^2+\rho_0\rho\cos\gamma}
       {\sqrt{f^2+\rho_0^2}\sqrt{f^2+\rho^2}}
\label{eq:sForLargef}\end{equation}
Here, the sign of $s_i$ depends on
whether the object is projected to the right(+) or left(-) of the line
from the image centre to $\vec{r}_0$'s projection.
If there is any image distortion, the distances $\rho_0$ and $\rho$ read
from the image have to be corrected accordingly.
Formula (\ref{eq:sForLargef}) is numerically inconvenient for large focal
lengths $f$ and small image distances $\rho$. For $\rho/f$, $\rho_0/f$
$\rightarrow$ 0 relation (\ref{eq:sForLargef}) can be approximated by 
\begin{equation}
  s_i^2 =(\frac{\rho_0}{f})^2+(\frac{\rho}{f})^2
        -2\frac{\rho_0\rho}{f^2}\cos\gamma
        + \mathcal{O}\big((\frac{\rho}{f})^4\big)
\label{eq:sForSmallf}\end{equation}
which is the law of cosines applied to the triangle in the image
plane in Fig.~\ref{fg:gets}. Hence, the angle $s$ can in this limit
be read directly from the image in units of arcsecs.

The necessary camera calibration information is then contained
in the position of the observers $\vec{r}_i$ and either the reference
point $\vec{r}_0$ or the two view directions $\vec{v}_i$ to the
reference point. In the latter case, $\vec{r}_0$ can be determined
from (see Fig.~\ref{fg:recon}a)
\begin{gather}
  \vec{r}_0=\vec{r}_1+d_1\vec{v}_1=\vec{r}_2+d_2\vec{v}_2\\
  \lvecc{\vec{v}_1}{\vec{v}_2}\rvecc{d_1}{-d_2} = \vec{r}_2-\vec{r}_1
\label{eq:refpoint}\end{gather}
Therefore a rectified image point $(s_1,s_2)$ = (0,0) will have to be
reconstructed at $\vec{r}_0$ and any other $(s_1,s_2)$ pair will
be conveniently expressed as a two-dimensional distance vector
$\Delta\vec{r}$ from $\vec{r}_0$.

For affine geometry, all view directions from an image $i$ are
approximated to be parallel to $\vec{v}_i$.
Then (see Fig.~\ref{fg:recon}b)
\begin{equation}
  \rvecc{\vec{e}_1^T}{\vec{e}_2^T} \Delta\vec{r} =
  \rvecc{d_1 \tan s_1}{d_2 \tan s_2}
\label{eq:affineDr}\end{equation}
If we subtract the two above lines we obtain the ``depth'' component of
$\Delta\vec{r}$ in direction half way between the two view directions.
Note, with $beta$ the angle between $\vec{v}_1$ and $\vec{v}_2$ we have
$\vec{v}_1+\vec{v}_2$ = $(\vec{e}_2-\vec{e}_1)/\tan(\beta/2)$. Then
\begin{equation}
   \tan\frac{\beta}{2}\; (\vec{v}_1^T+\vec{v}_2^T)\Delta\vec{r}
       = d_1 \tan s_1-d_2 \tan s_2
\label{eq:disparity}\end{equation}
where the difference on the right hand side is the {\bf disparity}.
In the simplest case, the depth of an object is directly proportional
to the disparity of its projections along the epipolar lines.

For projective geometry we take account of the divergence of the view
directions emerging from each observer. The angle $s_i$ between the
reference point $\vec{r}_0$ and an object in the epipolar plane then
is (see Fig.~\ref{fg:recon}a)
\begin{equation}
  \tan s_i=\frac{\vec{e}_i^T\Delta\vec{r}}
            {d_i+\vec{v_i}^T\Delta\vec{r}}
\nonumber\end{equation}
or
\begin{equation}
  \rvecc{\vec{e}_1^T-\vec{v}_1^T\tan s_1}
        {\vec{e}_2^T-\vec{v}_2^T\tan s_2}
  \Delta\vec{r} = \rvecc{d_1 \tan s_1}{d_2 \tan s_2}
\label{eq:prjgDr}\end{equation}
in contrast to (\ref{eq:affineDr}). To justify the more simple affine geometry
formula, $s_1$ and $s_2$ must both be small. For EUVI images, $s$ is of the order
of the the apparent solar disk radius of about 0.005 radian. 
For HI images, however, with a much larger field of view, projective geometry
is unavoidable. Even if $s$ is small some care is necessary, because
in (\ref{eq:affineDr}), the matrix on the left has to be inverted and small
changes in its parameters may alter its minor eigenvalue considerably.
The eigenvalue of a matrix $\lvecc{\vec{a}}{\vec{b}}$ is
$\mathcal{O}(\abs{\vec{a}\times\vec{b}})$ if it is small compared to
$\abs{\vec{a}}$ and $\abs{\vec{b}}$. Hence if the view directions are nearly
parallel, the difference between affine and projective geometry may be
non-negligible even if $s$ is small.

\begin{figure}
  \hspace*{\fill}
\includegraphics[bb=10 20 410 205, clip, width=8cm]{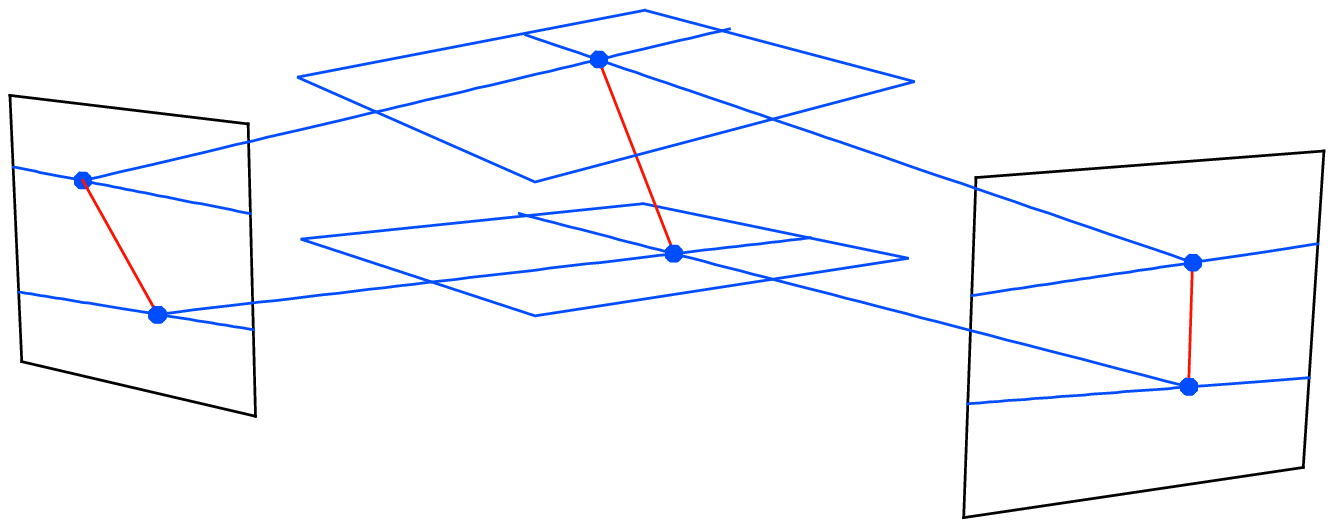}
  \hspace*{\fill}\\
  \hspace*{\fill}
\includegraphics[bb=10 20 410 205, clip, width=8cm]{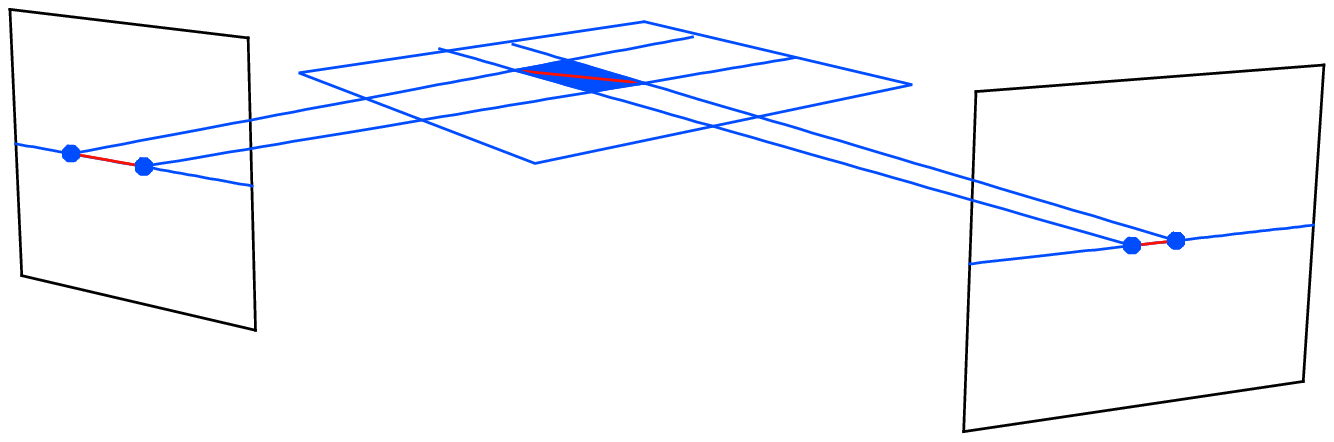}
  \hspace*{\fill}\\
\caption{Reconstruction of a curve segment (red) with different inclination
  with respect to the adjacent epipolar planes (blue). In the top drawing
  the line segment in inclined to, in the bottom drawing it lies exactly
  on an epipolar plane.
  (bottom).
\label{fg:curvsec}}
\end{figure}

For each epipolar plane we obtain a set of intersection points and all
we have to do is to connect the intersection points between
neighbouring planes to obtain three-dimensional curves.
This last step involves some uncertainties if different loops come
close. To obey the disparity continuity constraint we should keep
track of the intersections across the epipolar planes by
reconstructing curve segments between epipolar lines rather than
only points on single epipolar lines (Practically, we just combine the
intersection points of a given curve projection with epipolar lines
to ordered lists). Hence the method should better be called tie-curve
rather than tie-point method.

The curve segments from each image are then projected along their
respective geometrical view direction (either affine or projective)
which yields a narrow planar strip of the projection surface of the
curve (see Fig.~\ref{fg:curvsec}a).
The intersection of the strips from both images gives a small line segment
of finite length, the end points of which are two points of the
three-dimensional loop curve. The intersection is guaranteed since the
curve sections in the image were chosen from the same epipolar interval.

This strategy runs into a problem, however, if the curve segment is
directed parallel to an epipolar line (see Fig.~\ref{fg:curvsec}b). The
geometrical intersection of the two projection ribbons now is not
a line segment anymore but a small trapezoid. This object of dimension
2 is difficult to incorporate to our final three-dimensional loop
curve of dimension 1. This problem, however, is not a deficit of the
method but a fundamental geometrical limitation which will be
addressed in the following sections. It is a hint that the
reconstruction error at this point of the loop becomes singular and in
general, this is the point where the true solution for the loop curve
may intersects with ghost features.

\section{Reconstruction errors}\label{ch:errors}

\begin{figure}
\hspace*{\fill}
\includegraphics[width=10cm]{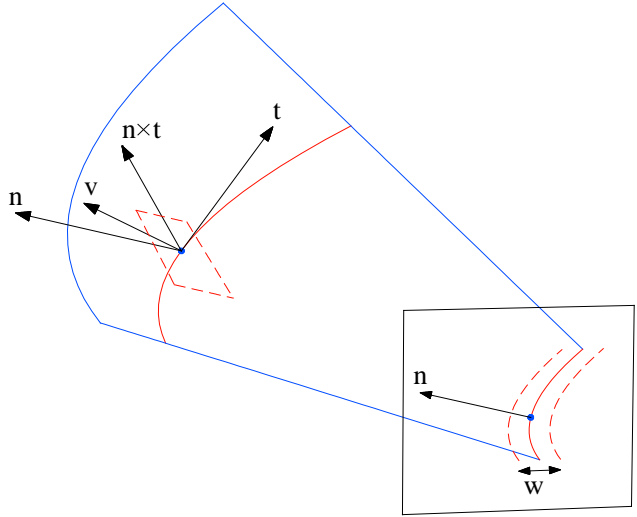}
\hspace*{\fill}\\
\caption{Local orthogonal coordinate system along the reconstructed
  loop (in red) used to express the uncertainty area (red dashed square)
  at a point along the loop.
  The position uncertainty in direction $\vec{n}$ is given by the
  resolution $w$ of the image.
  The precise position of the curve in direction $\vec{n}\times\vec{t}$
  can only be determined by the resolution of the second image.
  The projection surface of this second image is not shown for clarity,
  it intersects the projection surface from the first image (blue) along
  the red loop curve.
\label{fg:loopcoord}}
\end{figure}

If a loop's projection in an image can only be determined with a
finite resolution, there remains a positional uncertainty of the final
three-dimensional curve.
Consequently, we have to think of the projection surfaces as surfaces
of finite thickness and their intersection yields a tube with
trapezoidal cross section within which the final curve position cannot
be resolved.
The reconstruction uncertainty of a single point along the
three-dimensional curve due to the finite width $w_i$ of the loop in
image $i$ lies in the plane spanned by the local normals $\vec{n}_i$
of the two projection surfaces at the point where the projection
surfaces intersect (see Fig.~\ref{fg:loopcoord}).
Hence the positional uncertainty is distributed
normal to the local loop tangent
$\vec{t}=\vec{n}_1\times\vec{n}_2/|\vec{n}_1\times\vec{n}_2|$
as should be because uncertainties along the loop are irrelevant.

This error trapezoid at any point $\vec{r}$ along the loop can be expressed
as
\begin{equation}
  \delta\vec{r} =
  \pm a_1 \vec{t}\times\vec{n}_1 \pm a_2 \vec{t}\times\vec{n}_2
\label{eq:errvec}\end{equation}
where the $a_i$ are to be determined so that the projection of the
trapezoid onto image 1 and 2 yields the right loop width $w_1$ and
$w_2$, respectively. Upon multiplying (\ref{eq:errvec}) with $\vec{n}_1$
and $\vec{n}_2$ we obtain
\begin{equation}
  a_1=\frac{w_2}{2\sin\alpha} \quad\text{and}\quad a_2=\frac{w_1}{2\sin\alpha}
\label{eq:errcoeff}\end{equation}
where $\abs{\vec{n}_1\times\vec{n}_2}$ = $\sin\alpha$ and $\alpha$ is the
angle between the local projection surface normals.

\begin{figure}
\hspace*{\fill}
\includegraphics[bb=30 30 198 311, clip, width=5cm]{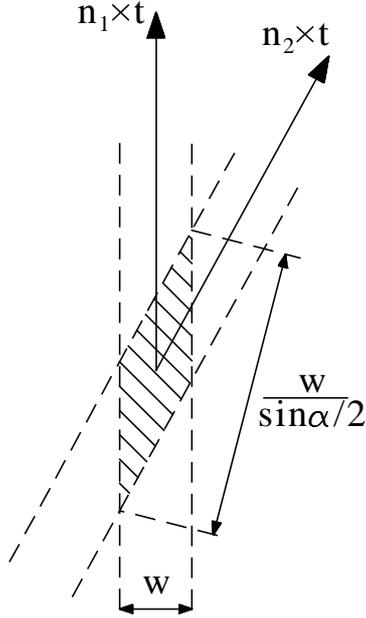}
\hspace*{\fill}\\
\caption{Error trapezoid in the plane normal to the loop. The loop's
  tangent $\vec{t}$ points out of the plane and $\alpha$ is the angle
  between $\vec{n}_1$ and $\vec{n}_2$.
\label{fg:errtrap}}
\end{figure}

If $w_1=w_2=w$ we have a local trapezoid with main axis length
(see Fig.~\ref{fg:errtrap})
\begin{equation}
  \frac{w}{2\sin\alpha} \vec{t}\times(\vec{n}_1\pm\vec{n}_2)
\label{eq:erraxes}\end{equation}
where the length of $(\vec{n}_1\pm\vec{n}_2)$ is
$\cos(\alpha/2)$ and $\sin(\alpha/2)$, respectively. The axis lengths
of the error trapezoid are therefore (measured from the centre to
the trapezoid corners)
\begin{equation}
  \frac{w}{2\cos(\frac{\alpha}{2})} \quad\text{and}\quad 
  \frac{w}{2\sin(\frac{\alpha}{2})}
  \label{eq:errsize}\end{equation}
We note, that the positional error with which we can reconstruct a loop
from finite resolution images does not depend on the stereo base angle
between the view directions $\vec{v}_i$ of the two observers but rather
on the angle $\alpha$ of the projection surfaces.

The two angles are not independent, though. $\alpha$ is always
smaller or at best equal to the stereo base angle. Where the loop tangent
$\vec{t}$ intersects an epipolar plane in normal direction, the
normals $\vec{n}_i$ are just the $\vec{v}_i$ rotated by 90 degrees.
In this case, $\alpha$ is identical to the stereo base angle. If, however,
the loop tangent lies in the intersected epipolar plane
(the situation sketched in Fig.~\ref{fg:curvsec}b), the
$\vec{n}_i$ are both the epipolar plane normal, hence are parallel
and $\alpha$ = 0 whatever the stereo base angle.

We have seen already in the previous section about the tie-point
reconstruction that we encounter a problem where the loop's images are
directed parallel to the epipolar lines. This complies with our error
estimate which becomes singular in this case.  

If, as at the beginning of the STEREO mission, the stereo base angle
between the two observing spacecrafts is small, (\ref{eq:errsize}) poses
a severe restriction to the resolution with which we can reconstruct
the loops. For a stereo base angle of 10 degrees, we have $\alpha$
$\le$ 10$^{\circ}$ and with $w$ set to the resolution of the STEREO
EUVI images of 2 arcsec, (\ref{eq:errsize}) gives $\pm$8000 km. Hence the
resolution in the direction half way between the two view directions
($\vec{n}_i\times\vec{t}$ is the view direction mapped to the plane
normal to $\vec{t}$) will not be better than about 16000 km in vertical
direction on the
solar surface. The resolution will, however, rapidly improve while the
STEREO spacecrafts separate. The error can also be reduced if the
position of the loop in the image can be determined with subpixel
precision by a suitable interpolation of the image intensities in the
vicinity of the loop projection.

In general, the numerically reconstructed loop curve will
stochastically vary within the above estimated error range due to
small fluctuations in the detected loop projection and numerical
round-off effects. Especially if the stereo base is small, the
resulting three dimensional curve may look very different from what we
expect to be a magnetic field line. A coronal magnetic field line should
show a smooth change of the local curvature vector $d\vec{t}/d\ell$
(see Frenet's formulas, $\vec{t}$ is the unit tangent and
$\ell$ the length coordinate along the loop). One way to improve
the tie-point method would be to not only reconstruct the curve, but
along with it store the local error axes and their length estimates
for every loop point.
In a post-processing step the three-dimensional curve is then smoothed
(e.g., by a smoothing spline) within these error bounds so that the
variation of the curvature vector along the loop is minimized.
In its extremes, this approach leads to loop curves with constant
curvature, i.e., sections of circles as proposed by
\citet{AschwandenEtal:1999}.

\begin{figure}
\hspace*{\fill}
\includegraphics[bb=117 43 508 799, clip, height=\textwidth, angle=-90]
                {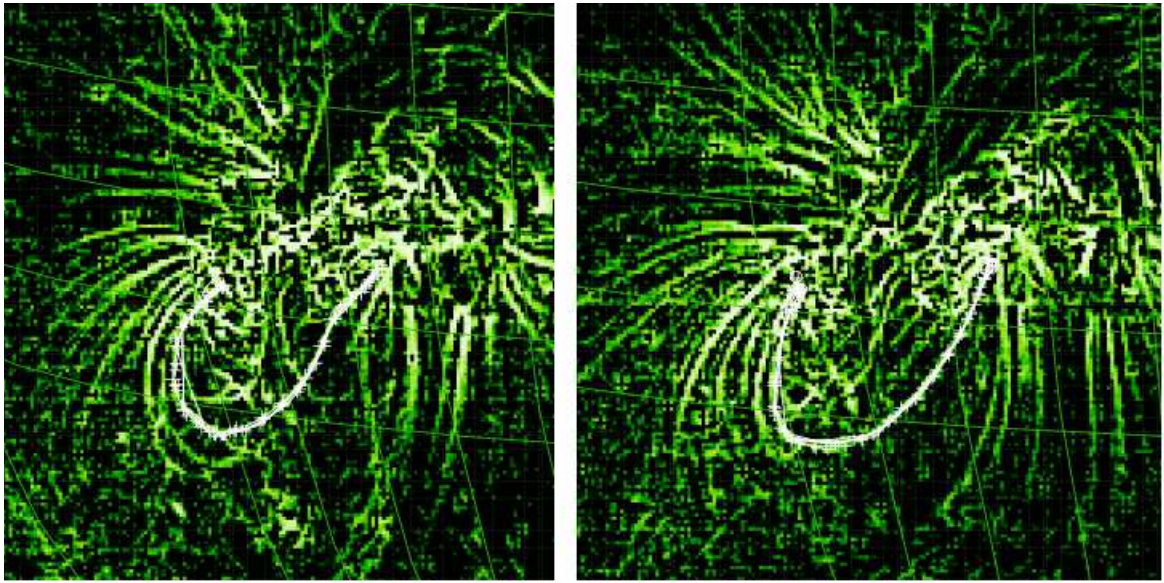}
\hspace*{\fill}
\caption{EIT observations of an active region taken 23 hours apart
  In white we show the attempts to trace out a loop which is well
  visible in the contrast enhanced image.
\label{fg:eit2}}
\end{figure}

\begin{figure}
\hspace*{\fill}
\includegraphics[bb=45 52 339 345, clip, width=6cm]{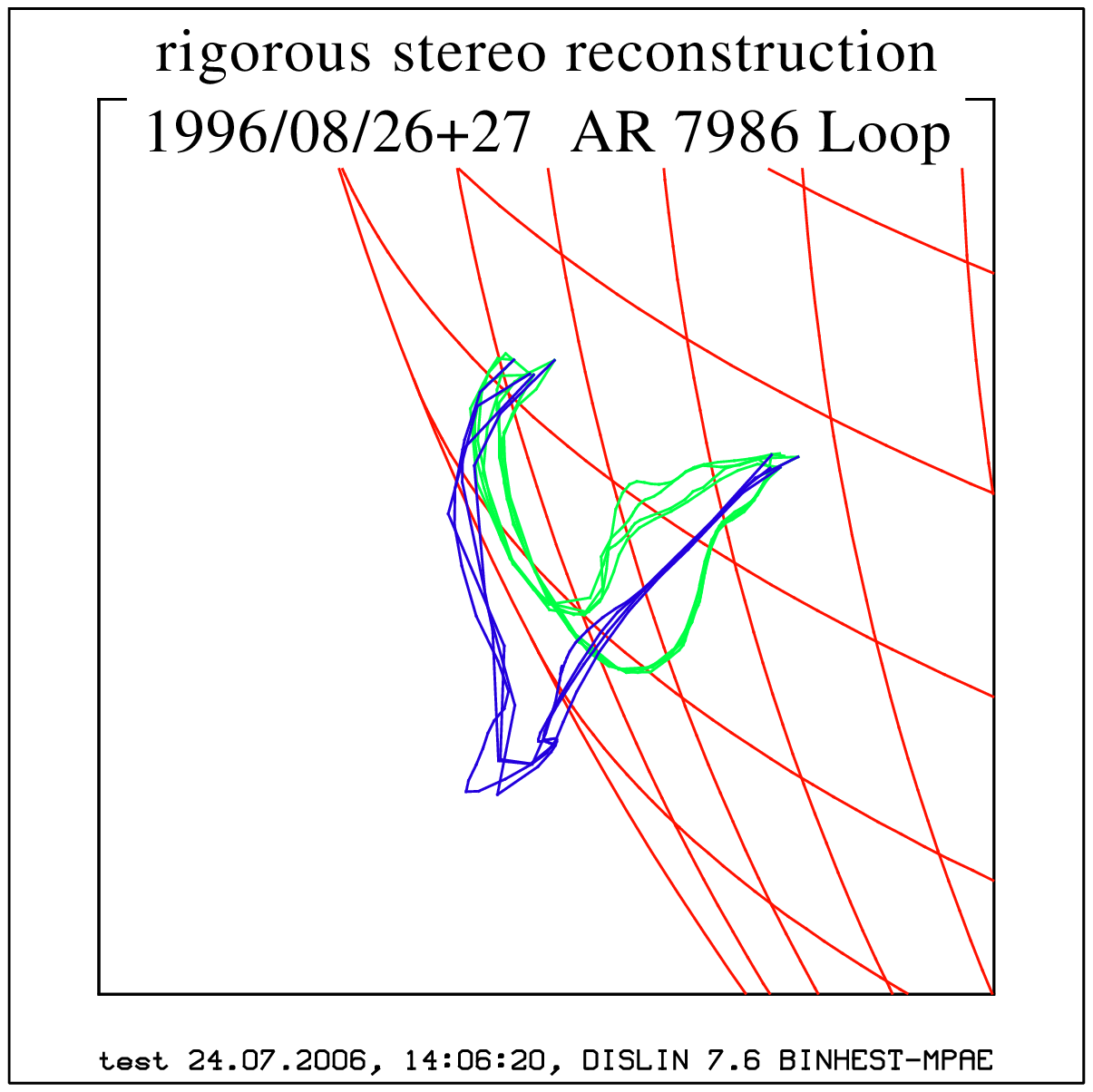}
\includegraphics[bb=45 52 339 345, clip, width=6cm]{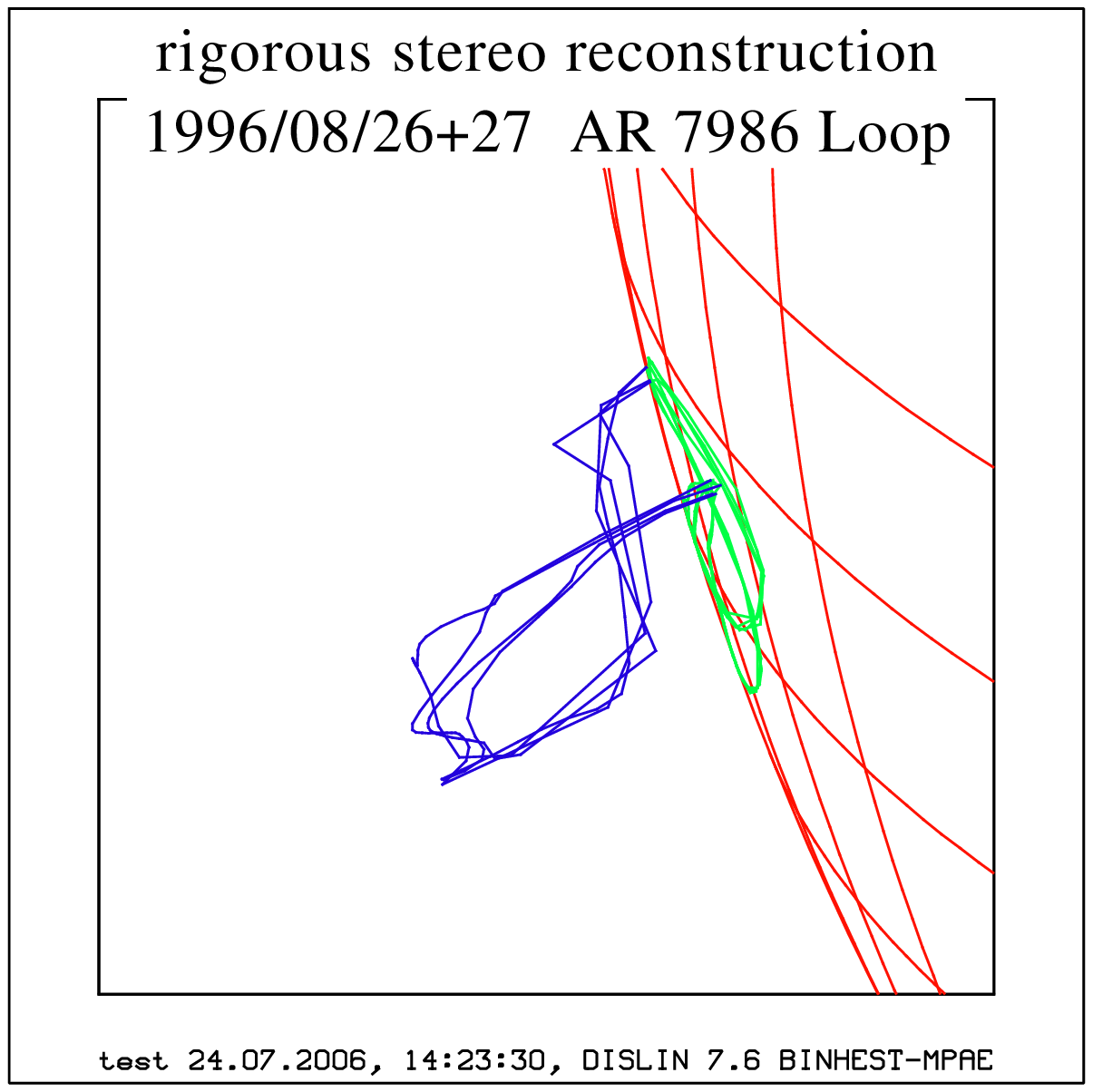}
\hspace*{\fill}
\caption{Results of the reconstruction (blue) from the loop projections
  shown in Fig.~\ref{fg:eit2} viewed at from two different directions
  which do not coincide with the view directions of the observations. 
  The solar surface is indicted by a latitude/longitude grid (red) and
  the loop projections are mapped as green curves onto the solar surface.
\label{fg:stercon7a}}
\end{figure}

As an example, we show the reconstruction attempts for a single
field line of an active region from a pair of EIT observations
in Figs.~\ref{fg:eit2} and \ref{fg:stercon7a}.
Due to the present lack of true STEREO data, two images were
selected which were taken approximately 23 hours apart. The
solar rotation then produces an equivalent stereo base angle of 12.8
degrees. To make the loop structures better visible, the images were
contrast enhanced with an filter of even order (Note that filters of
odd order such as Sobel are less noisy but may shift the intensity
enhancement away from the loop centre).

Fig.~\ref{fg:eit2} displays the manually traced loop projections on
top of the respective EIT image while Fig.~\ref{fg:stercon7a} shows
the reconstruction from the loop projections. The reconstructed loops
are displayed from two view points, each different from the observing
directions.
The second of these views which shows the reconstructed loop just
above the limb reveals the substantial variation of the reconstructed
curve in vertical direction, approximately the view directions of the
two observations. This insensitivity of the reconstruction in height
is a consequence of the small stereo base angle as discussed above.
Besides this large vertical error, the loop very probably also has slightly
changed its shape in the time interval between the observations of
image 1 and 2 which might also add to its odd appearance. This can
somewhat be guessed from the kinks in the loop projection of image 1
which are absent in image 2.

\begin{figure}[t]
\hspace*{\fill}
\includegraphics[bb=45 52 339 345, clip, width=6cm]{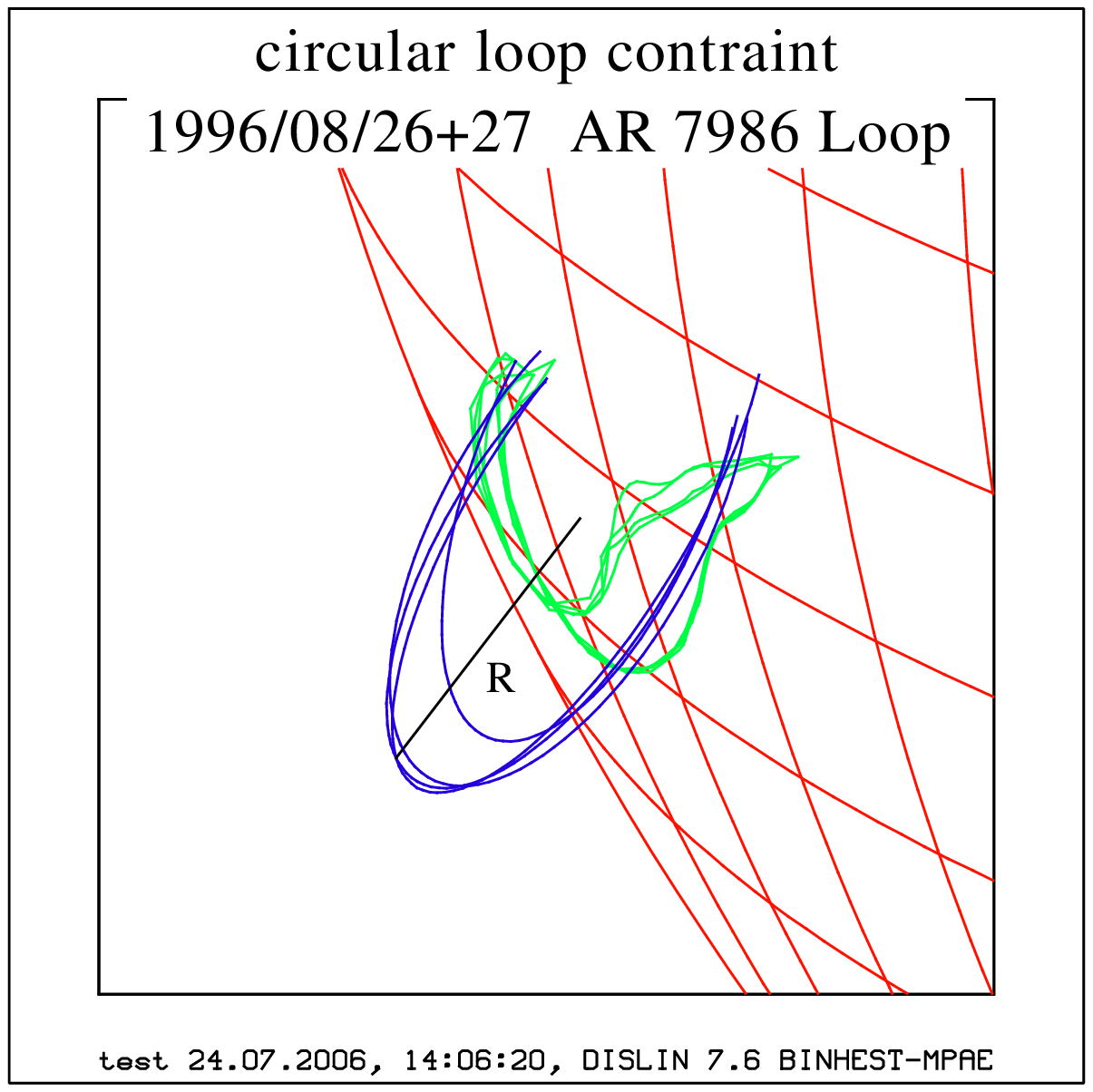}
\includegraphics[bb=45 52 339 345, clip, width=6cm]{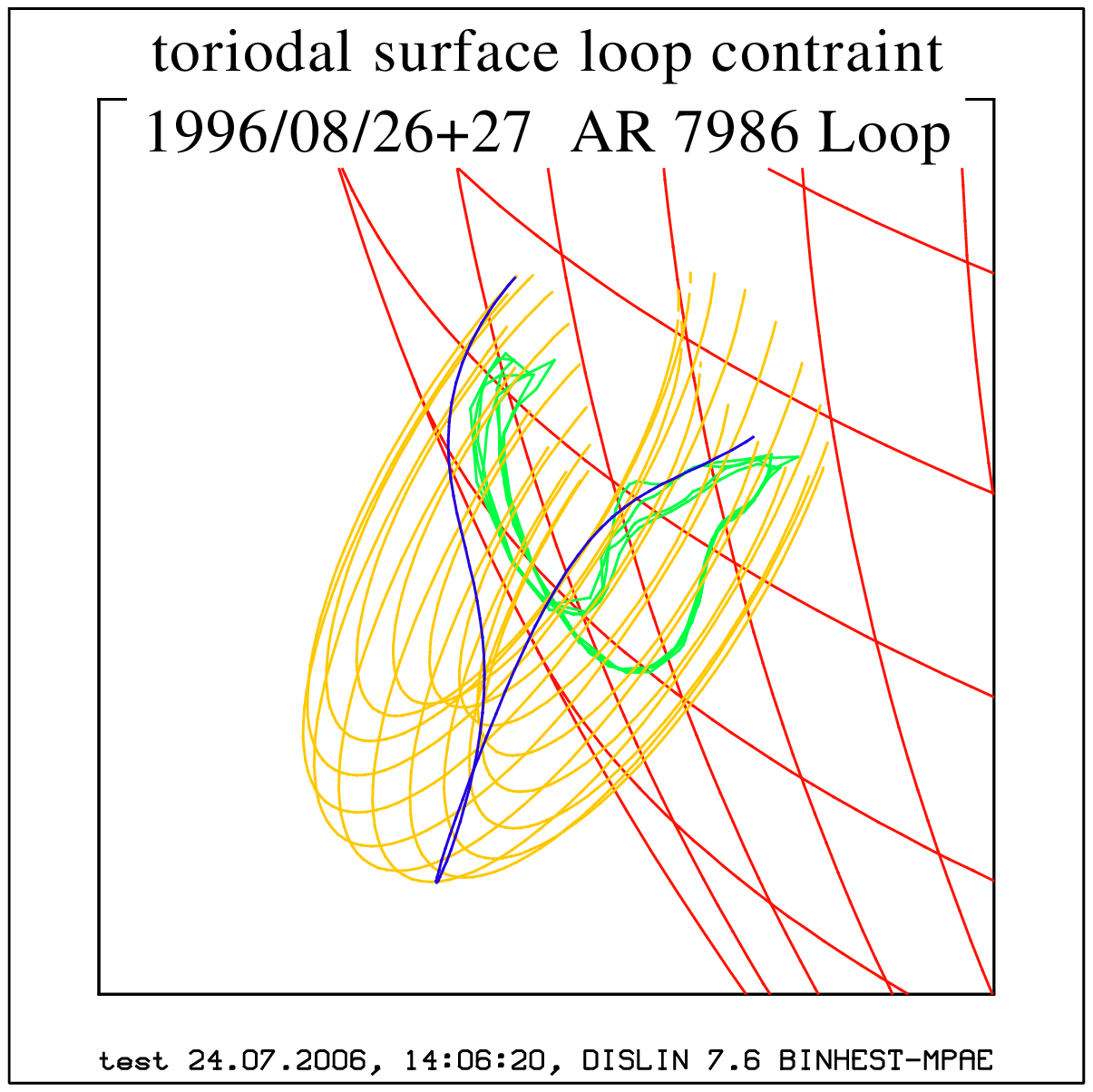}
\hspace*{\fill}
\caption{A circle and a torus fitted to the observations.
\label{fg:stercon7b}}
\end{figure}

Finally, Fig.~\ref{fg:stercon7b} shows the attempt to force the loop
into a shape with a small variation of the curvature vector. The loop
was fitted to a circle of constant radius (left) and to the surface of a
torus of constant major and minor radii (right). These approaches have
been suggested by \citet{AschwandenEtal:1999,Aschwanden:2005}.

From the forthcoming STEREO mission, we hope to obtain image data
which has better resolution than the EIT data so that such an extreme
smoothing of the reconstructed curves will not be necessary.

\section{Ambiguities}\label{ch:ambiguities}

One source of ambiguous reconstruction results is our limited ability
to make definite correspondences between loop projections. Every
alternative correspondence produces a new individual loop or possibly
even a whole set of new loops. If the constraints listed in
section~\ref{ch:matching}) do not help us to discard multiple
correspondences, the image data alone cannot be interpreted in an
unambiguous way.

\begin{figure}
\hspace*{\fill}
\includegraphics[width=10cm]{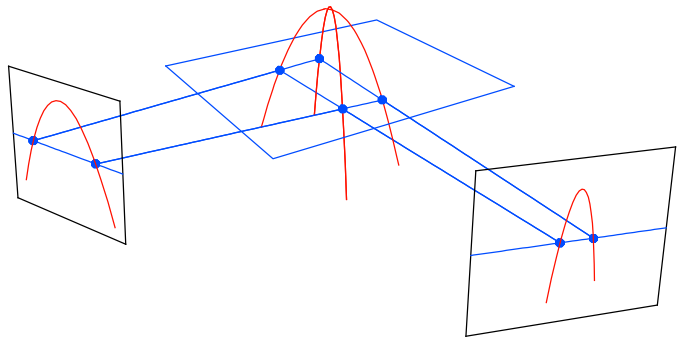}
\hspace*{\fill}\\
\caption{Double solutions obtained from the reconstruction of a
  single arc. In each image, the projection covers 
  a certain epipolar range twice with a turning point in between. 
\label{fg:ambig2}}
\end{figure}

Even individual loops can fool us. The simplest example is a magnetic
arc (Fig.~\ref{fg:ambig2}) which, if viewed from two different directions
can be interpreted in two ways represented by the two crossing arcs.
A typical feature of these ambiguous reconstructions is that they
intersect exactly at an epipolar turning point where the loop projections
become parallel to the epipolar line. We recall that this is a critical
point in the reconstruction  anyway because the reconstruction error
becomes singular at this point.

\begin{figure}
\hspace*{\fill}
\includegraphics[width=10cm]{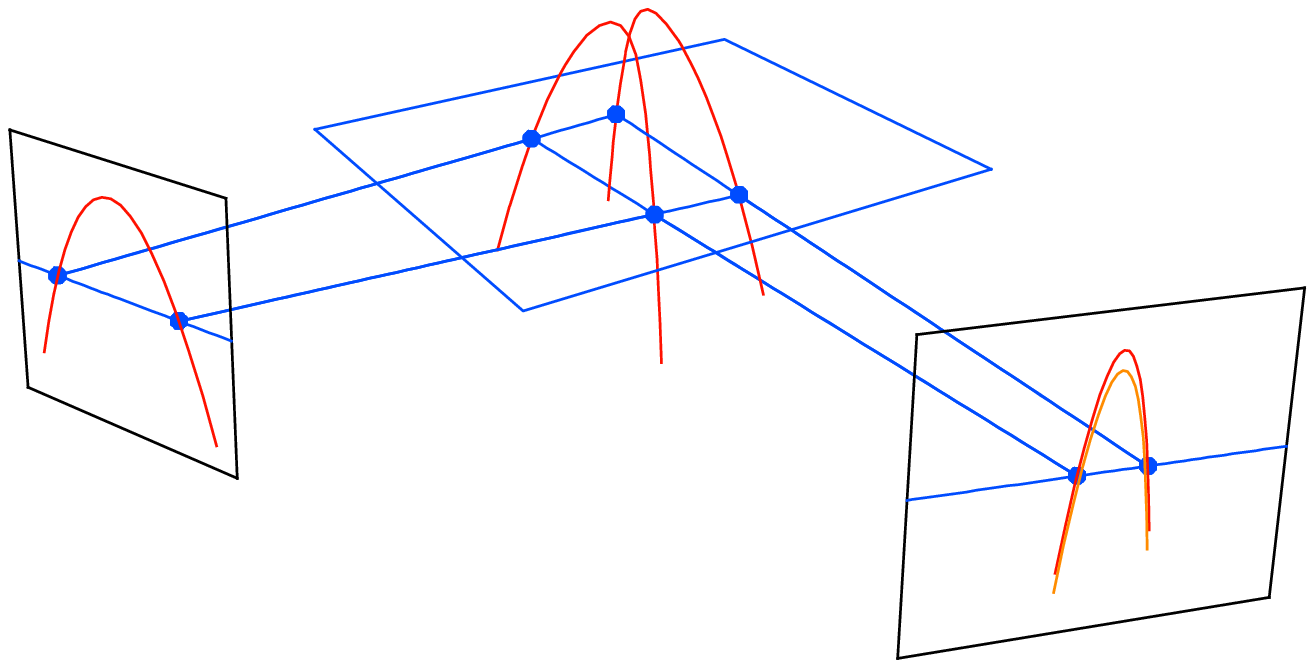}
\hspace*{\fill}\\[-5mm]
\hspace*{\fill}
\includegraphics[width=10cm]{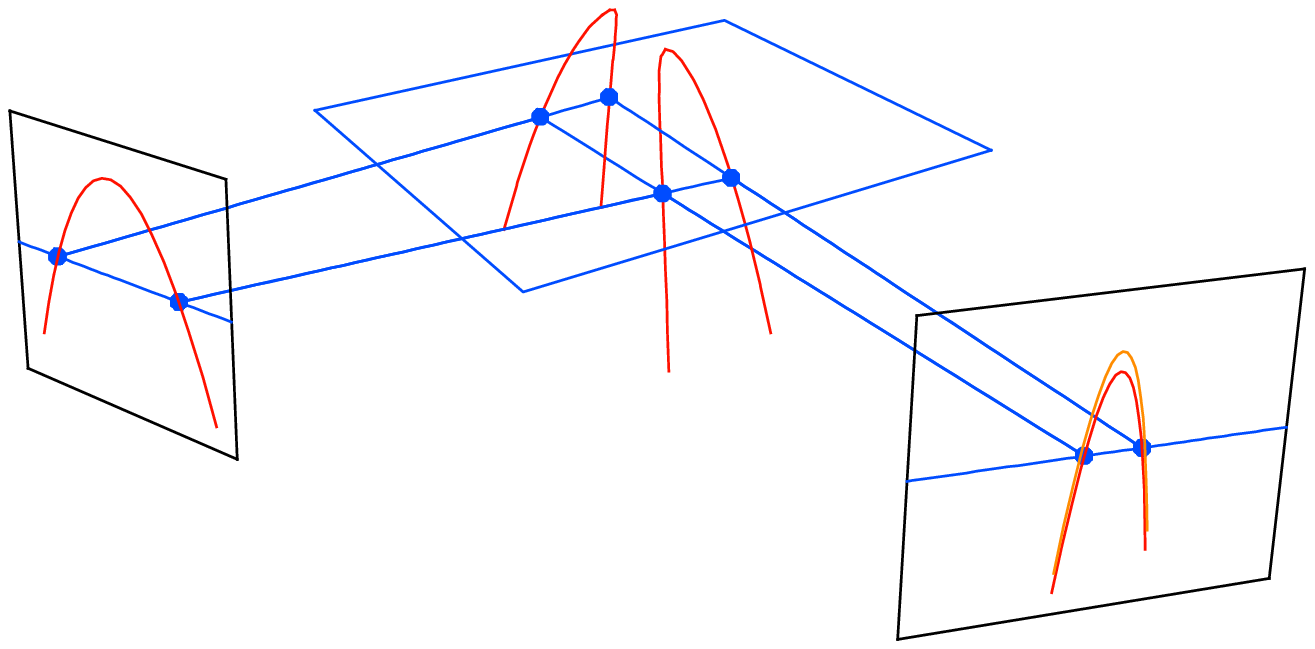}
\hspace*{\fill}\\[-5mm]
\caption{Apparent field line reconnection obtained from the
  reconstruction of a single magnetic arc where one of the projections
  was erroneously shifted upwards (top) and downwards (bottom).
  The projection in light red in the right image indicates the
  original position.
\label{fg:ambig2a}}
\end{figure}

This is in particular demonstrated if we add little errors to one or both of
the loop projections as in Fig.~\ref{fg:ambig2a}. Here, the
projection in the right image was shifted up- and downwards by a small
amount. The crossing arcs split in a characteristic way with the
deviations from the original solution being biggest at the top of the
arc. Both observations if made in sequence would appear as if we had
observed field line reconnection and yet it's only a combination of
finite measurement errors and stereoscopic ambiguity.

In the following we will define the reconstructed curve sections, which
due to small errors may connect with each other in different ways as
{\bf curve branches}. Since the reconnections obviously occur at the
epipolar turning points, each branch consists of a curve section with
a monotonously varying epipolar coordinate along the curve.

With a finite image resolution included, it may well occur, that a
branch of the true reconstruction solution combines smoothly with a
ghost branch. This phenomenon may make a distinction between true and
ghost branches particularly difficult. These false connections,
however, can only occur where the loop projections are directed
parallel to the epipolar lines. This condition may help to concentrate
post-processing checks of the reconstructed loops on these critical
points.

Common to the above examples in Figs.~\ref{fg:ambig2} and \ref{fg:ambig2a}
is that the loop projections in both images cover a certain epipolar range
twice with the critical epipolar turning point right in between.
We find from the reconstruction four
branches each monotonous in the epipolar coordinate.
This principle can fairly easily be generalized: a loop projection,
which crosses a certain epipolar line $n$ times should only match with
a projection in the other image which also intersects this 
epipolar line $n$ times (intersections which fall on top of each other
are counted with according multiplicity).
If this extends over a finite epipolar range, the rigorous
reconstruction yields $n\times n$ branches, some of which may connect
at the boundaries of the epipolar range if a turning point in the loop
projection is present at the respective epipolar line.

\begin{figure}
\hspace*{\fill}
\includegraphics[width=10cm]{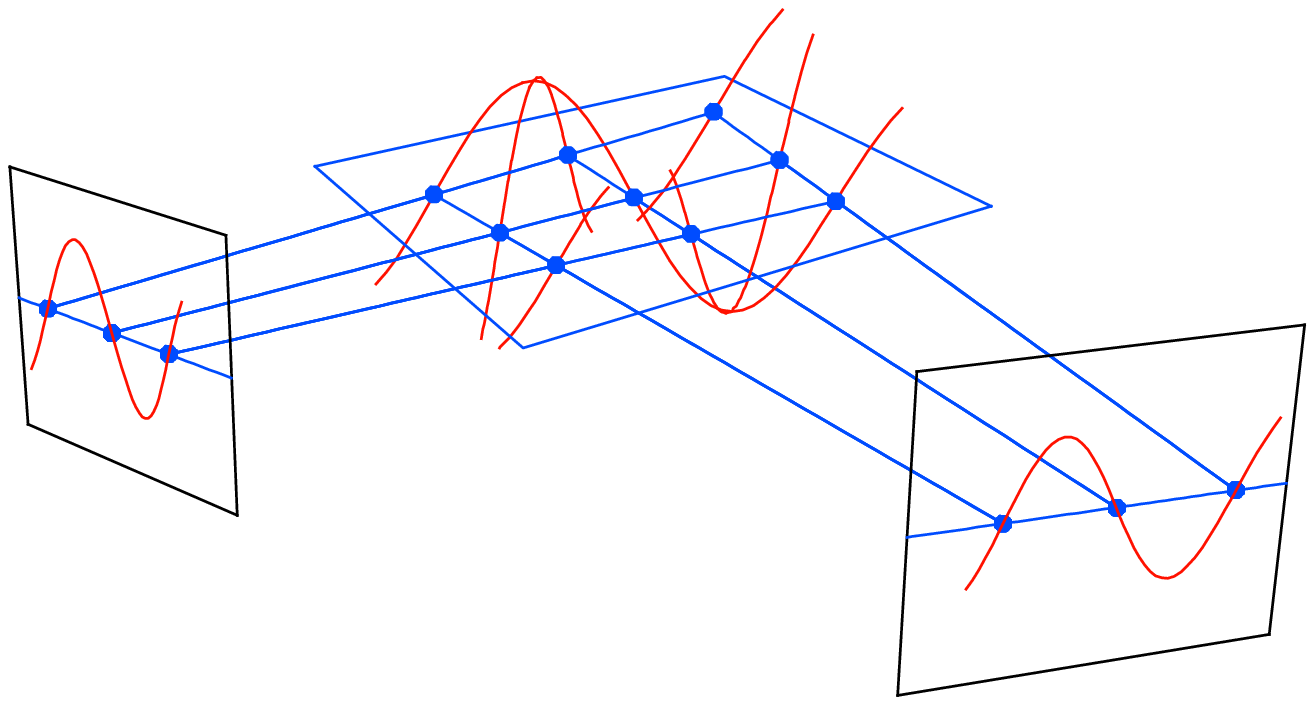}
\hspace*{\fill}\\[-5mm]
\hspace*{\fill}
\includegraphics[width=10cm]{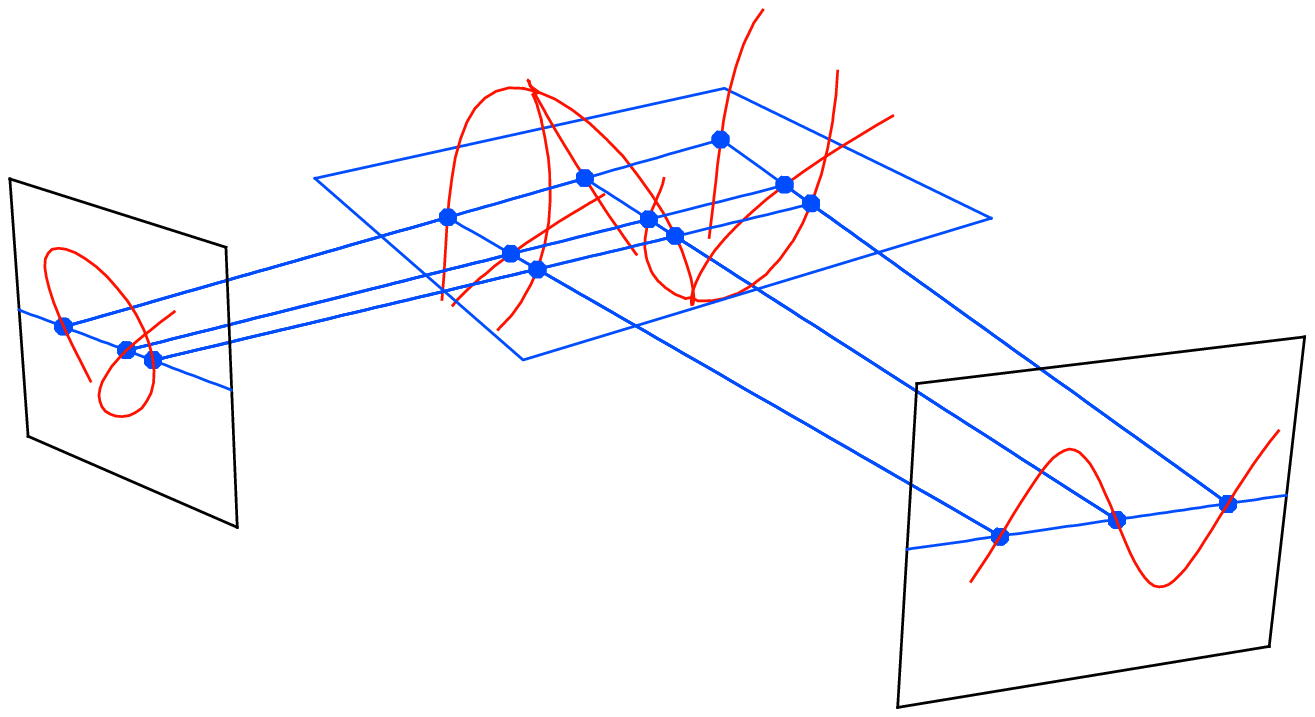}
\hspace*{\fill}\\[-5mm]
\caption{Multiple solutions for the reconstruction of a loop
  which covers a certain epipolar range three times. 
\label{fg:ambig3}}
\end{figure}

We demonstrate this principle in Fig.~\ref{fg:ambig3} with a
threefold coverage of a certain epipolar range by the loop projections.
We see that nine branches can be reconstructed, each monotonous in the
epipolar coordinate. Two of these remain isolated, four more connect
pairwise and only the remaining four connect to a continuous curve which
produces the full loop projection. Hence if we expect the true solution
to form a single continuous loop, this last one has to be the right
solution and all other branches are ghost features.

\begin{figure}
\hspace*{\fill}
\includegraphics[width=10cm]{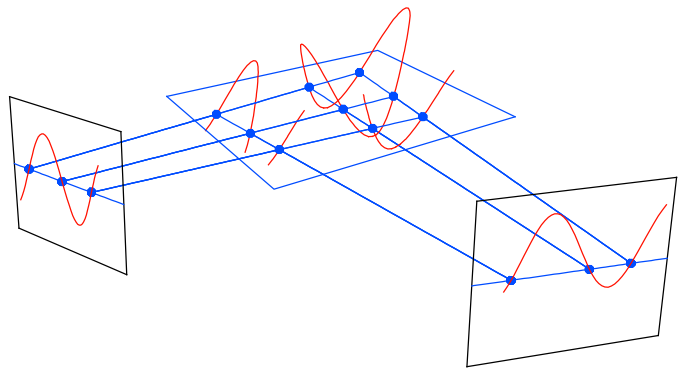}
\hspace*{\fill}
\caption{Typical closed loop ghost features which result if the
  curves projections in each image has extrema at different
  epipolar values.
\label{fg:ambig5}}
\end{figure}

So far we have assumed that we have found the right pair of
projections originating from one loop. We may erroneously make a wrong
correspondence and try to reconstruct from projections which do not
belong to the same original loop. Typical reconstruction branches which may
result from such an attempt are shown in Fig.~\ref{fg:ambig5}. We see
isolated branches and branches connected at both ends so that they
form a closed curve. None of the connected branches covers the entire
projection in both images. This may be taken as a hint that the two
projections do not make a valid correspondence.

\begin{figure}
  \hspace*{\fill}
\includegraphics[width=12cm]{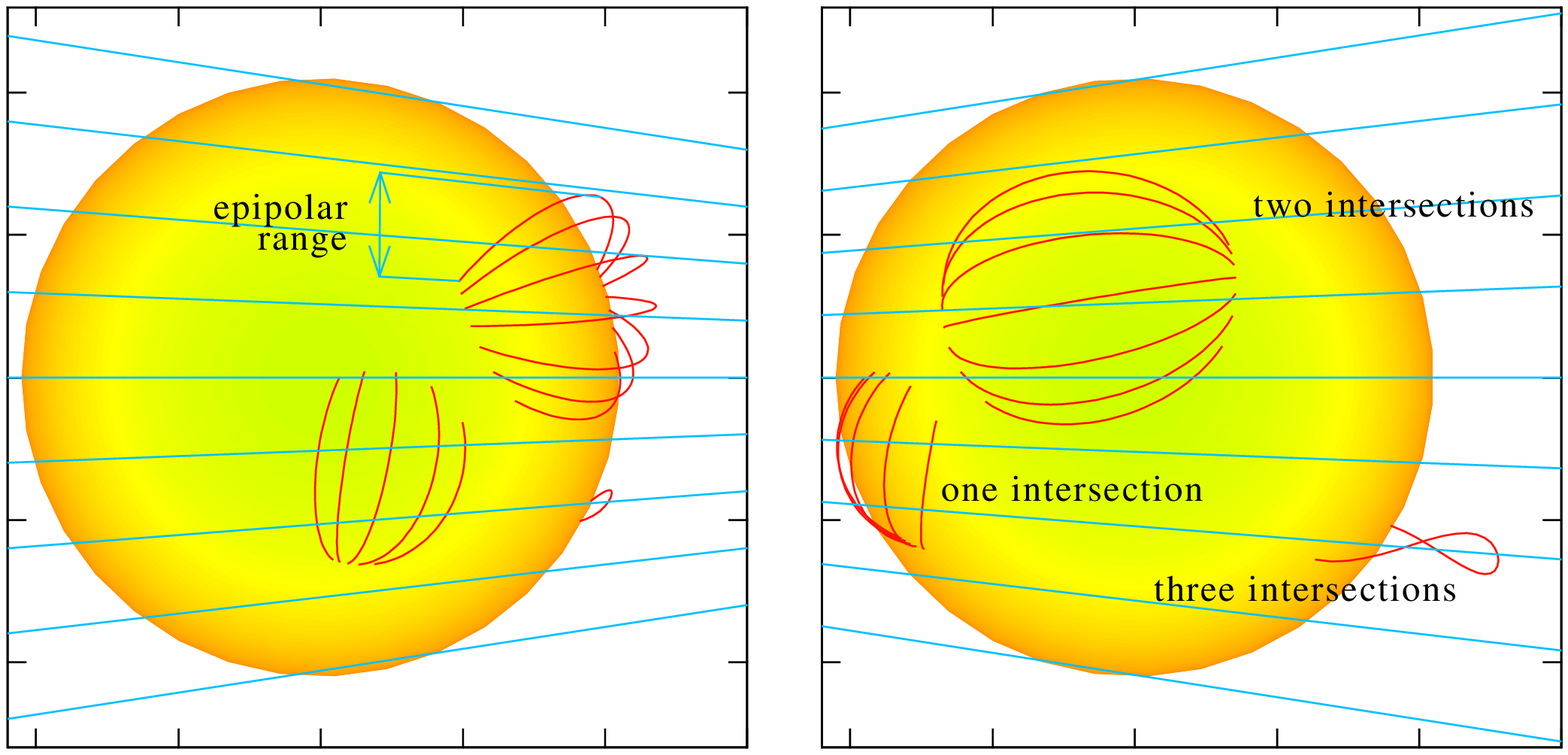}
  \hspace*{\fill}
\caption{A stereo image pair with groups of loops which have a different
number of intersections with epipolar lines. Conjugate pairs of epipolar 
lines are superimposed on each image.
\label{fg:ambigexmpl}}
\vspace{10mm}
\hspace*{\fill}
\includegraphics[bb=30 30 530 350,clip, width=12cm]{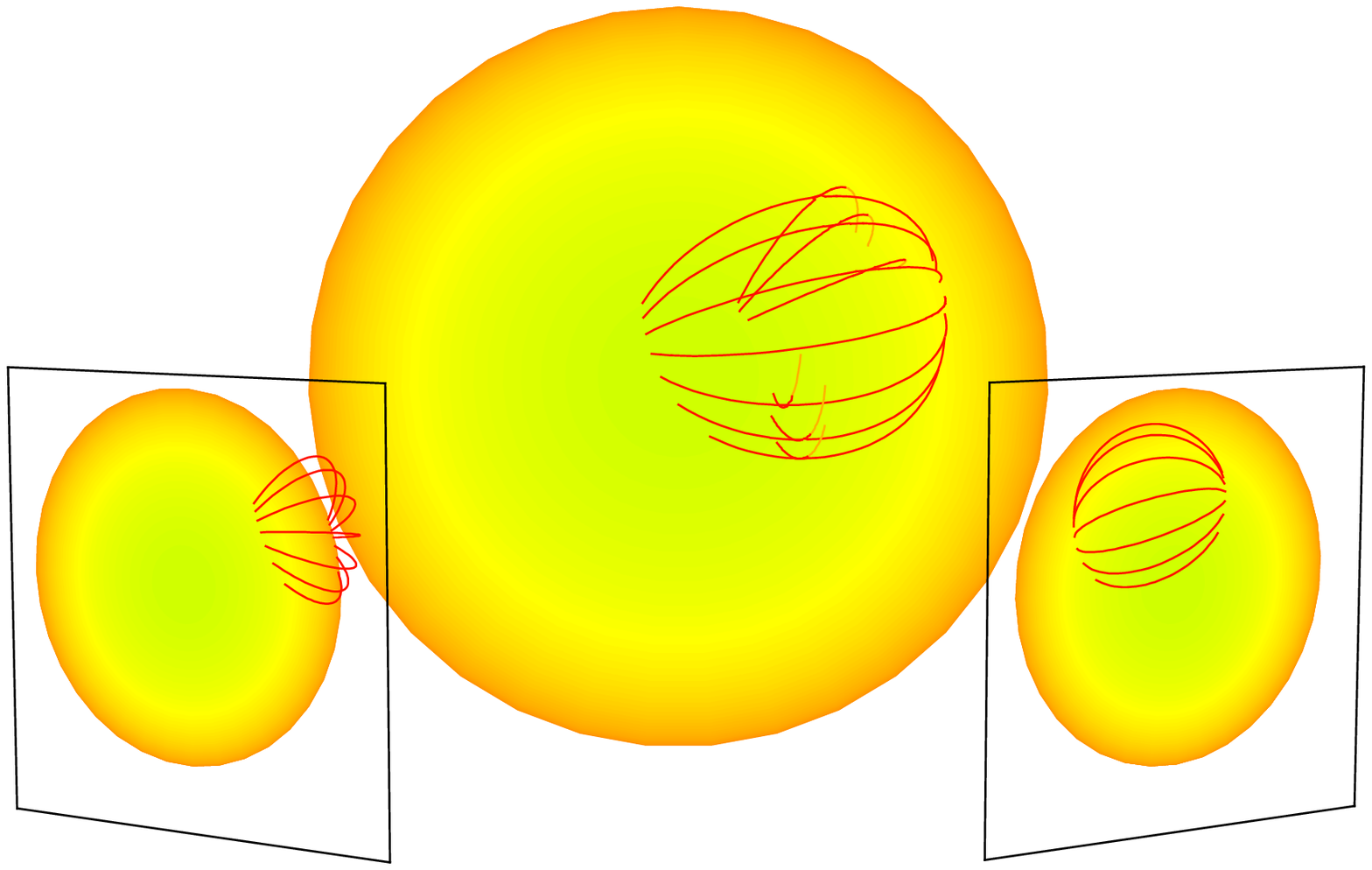}
\hspace*{\fill}\\
\caption{Reconstruction for one of the group of loops of  
Fig.~\ref{fg:ambigexmpl}. The reconstructed curves are
drawn in red, except for those parts which
lie inside the Sun and which are drawn in light red.
\label{fg:ambigexmplb}}
\end{figure}

Fig.~\ref{fg:ambigexmpl} summarizes our findings. Since
epipolar lines for the STEREO mission are only slightly inclined with
respect to the ecliptic, loops in north-south direction are straight
forward to reconstruct once they have been identified in both images.
An example is the arcade on the southern hemisphere in
Fig.~\ref{fg:ambigexmpl}.

Loops with east-west orientation usually intersect some epipolar lines
twice if they are curved. The example here is the set of loops in the
northern hemisphere in Fig.~\ref{fg:ambigexmpl}. The only
exception is the loop at the centre of this group. We have to expect
therefore that the reconstruction of these loops is ambiguous and
yields two crossing arcs each. Since the individual loops do not cover
a large epipolar range we can even anticipate that the pair of
reconstructed arcs makes only a small angle with each other and may be
difficult to distinguish. Fig.~\ref{fg:ambigexmpl} shows the result
of the reconstruction of these loops.
In fact, for the reconstruction of the central loop which is the only
unique one, the pair of arcs coalesces to a single arc.
Unfortunately these more complicated east-west orientated loops are
far more common for active regions on the solar surface.

Less common and even harder to reconstruct are loops with triple
crossings of an epipolar line as indicated by the individual loop
above the southern limb in the right image. In this case a
reconstruction is impossible anyway because the loop is hidden behind
the solar limb in the left image.

\section{More than two viewpoints}

We learned above that the epipolar geometry is especially adapted
to an image pair. A third or even more images are difficult to include
directly unless certain simplifying assumptions concerning the overall
geometry are made. The simplest way to make use of additional
(simultaneous) images is to produce stereo reconstructions from pairwise
combinations of the images. Since the ghost features very probably will
be located differently in the reconstructions from different image
combinations, the intersection of all reconstruction models should eliminate
most of the ghosts.

\begin{figure}[t]
\hspace*{\fill}
\includegraphics[bb=80 60 500 350, clip, width=10cm]
                {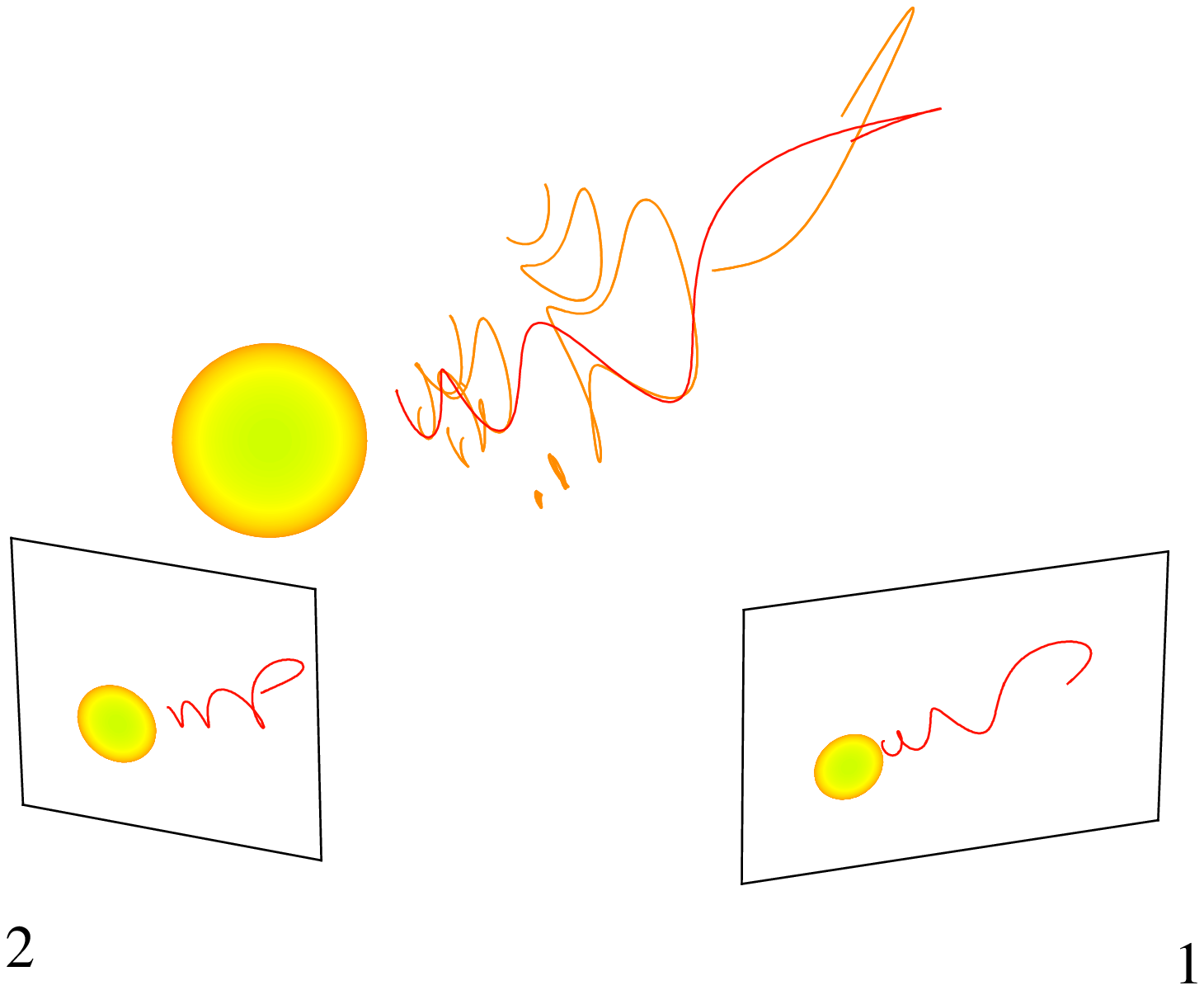}
\hspace*{\fill}\\
\hspace*{\fill}
\includegraphics[bb=80 60 500 350, clip, width=10cm]
                {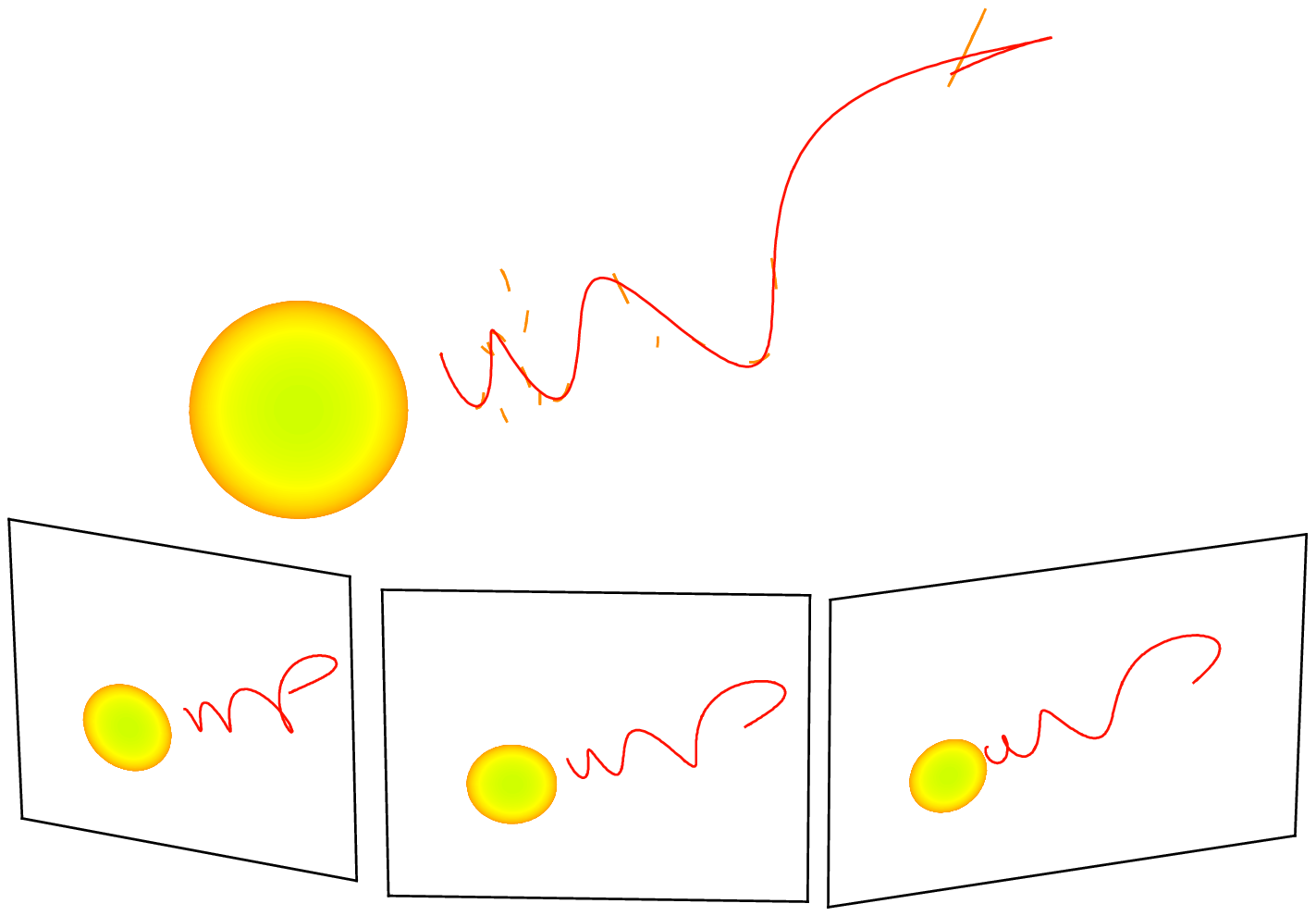}
\hspace*{\fill}
\caption{Stereo reconstructions of a CME-like extended flux rope
  (red) including ghost features (light red). The top panel shows
  the reconstruction based on two stereo images, the bottom panel
  the reconstruction from three simultaneous images taken all from the
  ecliptic plane.
\label{fg:cme3}}
\end{figure}

In order to demonstrate the advantage of a third view we consider a
more extended and more complex object as above. Fig.~\ref{fg:cme3}
shows a very simplified model of a CME flux-rope. The original field
line curve is drawn in red and it is faithfully reconstructed. However
along with it the reconstruction includes a number of ghost features
drawn in light red in the top of Fig.~\ref{fg:cme3}. We recognize the
isolated curve branches and closed field line curves typical for ghost
features as in Fig.~\ref{fg:ambig5}. In the magnetic cloud of a CME
closed field lines of such shape very probably do not occur.

If we include a third view in the reconstruction, e.g., from a
near-earth position, we can produce three reconstruction models which,
if properly done, all include the true field line curve. The ghost
features from the different reconstructions on the other hand hardly
overlap and can thus be almost completely eliminated except for a few
short line segments which are easily recognized as false in the bottom
diagram of Fig.~\ref{fg:cme3}.

For the STEREO mission, the ``third eye'' could be provided by ground
based coronagraphs or by the SOHO/EIT and LASCO instruments. All three
observers would then be placed in the ecliptic at almost equal
angular distance to its neighbour. An even more advantageous
arrangement would be a third spacecraft out of the ecliptic. In this
case the mutual epipolar plane systems would make a large angle with
respect to each other. In this geometry, critical reconstruction
points where loop tangents are parallel to the epipolar lines in one
epipolar system would be regular reconstruction points in the other
systems and reconstruction errors could be effectively reduced.
For the arrangement as shown in the lower part Fig.~\ref{fg:cme3},
the critical reconstruction points stay close together in all three
epipolar systems and the error prone loop sections remain. 

The advantage of a third view also only pays if the stereo base
angle is large enough. Recalling our error estimates in section
\ref{ch:errors} the mutual angles between the view directions should
be larger than about 10 degrees. Otherwise the geometrical uncertainty 
may partly be larger than the distance between true and ghost
curve features.

\section{Magnetic field}\label{ch:magnetic}

The fact that the plasma striations we want to reconstruct are strictly
aligned along the magnetic field suggests that we use magnetic field
information in addition to the image data to help with the reconstruction.

\begin{figure}[t]
\hspace*{\fill}
\includegraphics[bb=95 365 500 750, clip, height=10cm]{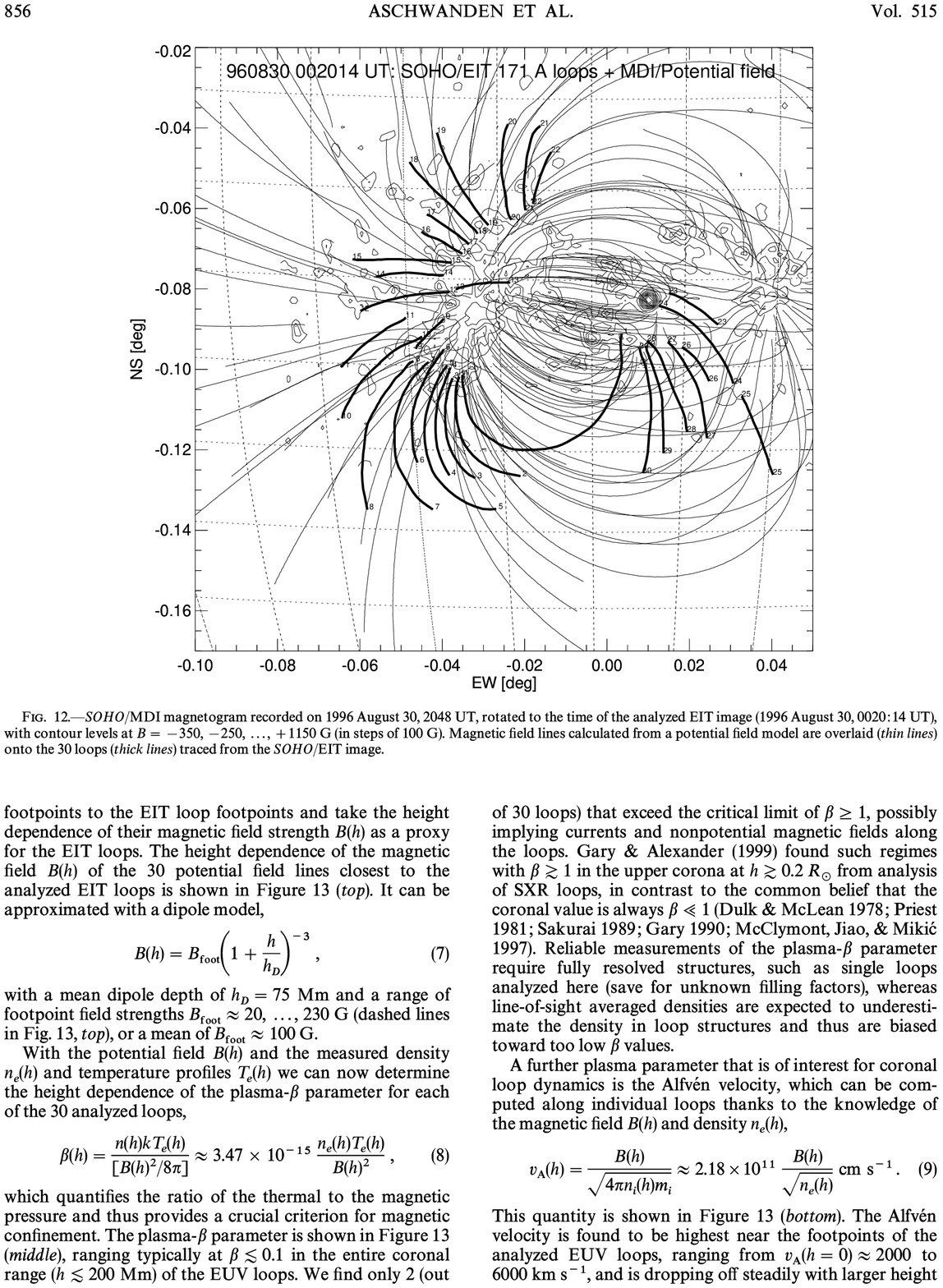}
\hspace*{\fill}
\caption{Observed loop projections (thick) and projected potential
  field lines (thin) for an active region with a surface normal
  magnetic field component indicated by the thin contour lines.
  From \citet{AschwandenEtal:1999}.
  \label{fg:aschm}}
\end{figure}

Magnetic field observations are available from solar surface Zeeman
and Hanle effect observations. Hardly any direct field measurements
have been made in the corona because line intensities there are much
fainter, thermal line spread much larger and field magnitudes much
less than in the photos- and chromosphere. The coronal field is
therefore often deduced from the surface observations by means of
field extrapolation. The extrapolation requires assumptions about the
the complexity of the field. The simplest and most often used is the
assumption that electric currents are absent and that the magnetic field
is a Laplace field. Observations show, however, that field lines
calculated under this assumption often deviate from observed loops,
especially near active regions (see, e.g., \citet{AschwandenEtal:1999}
from where we took Fig.~\ref{fg:aschm} and \citet{WiegelmannEtal:2005}).

\begin{figure}
\hspace*{\fill}
\includegraphics[bb=160 90 260 340,clip,height=9cm,width=6cm]{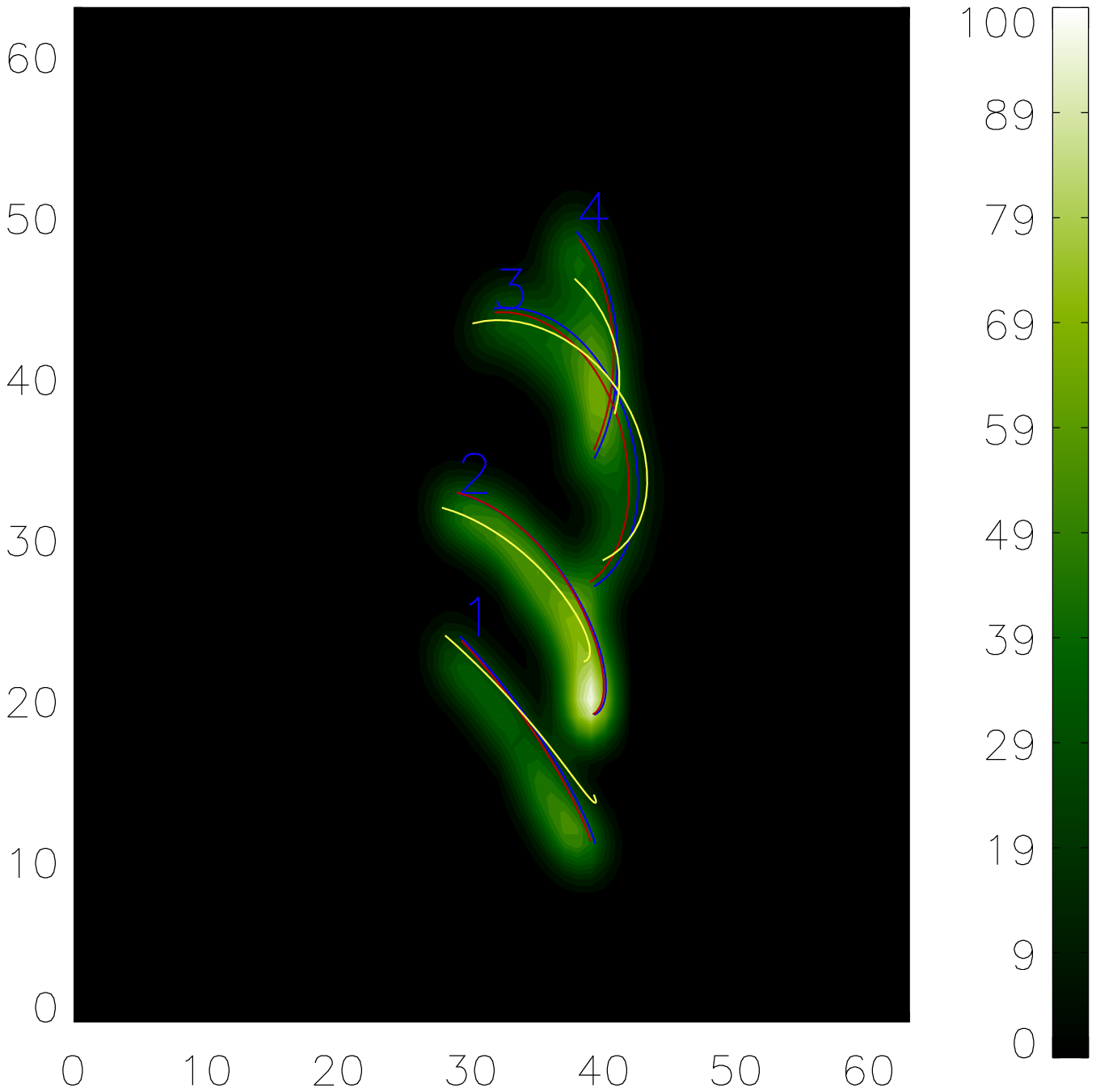}
\hspace{5mm}
\includegraphics[bb=160 90 260 340,clip,height=9cm,width=6cm]{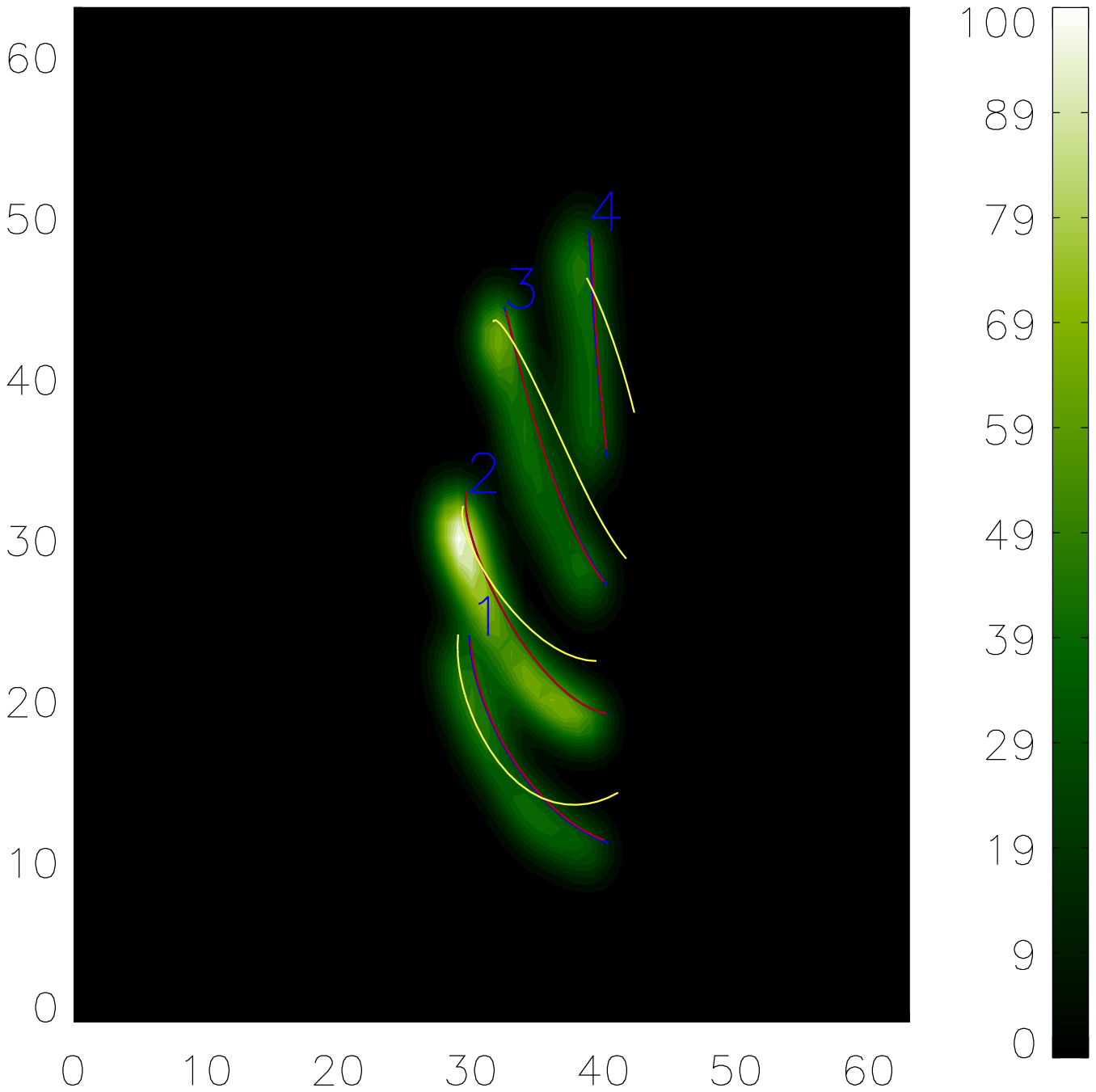}
\hspace*{\fill}\\[5mm]
\hspace*{\fill}
\includegraphics[bb=190 50 300 190,clip,height=6cm,width=6cm]{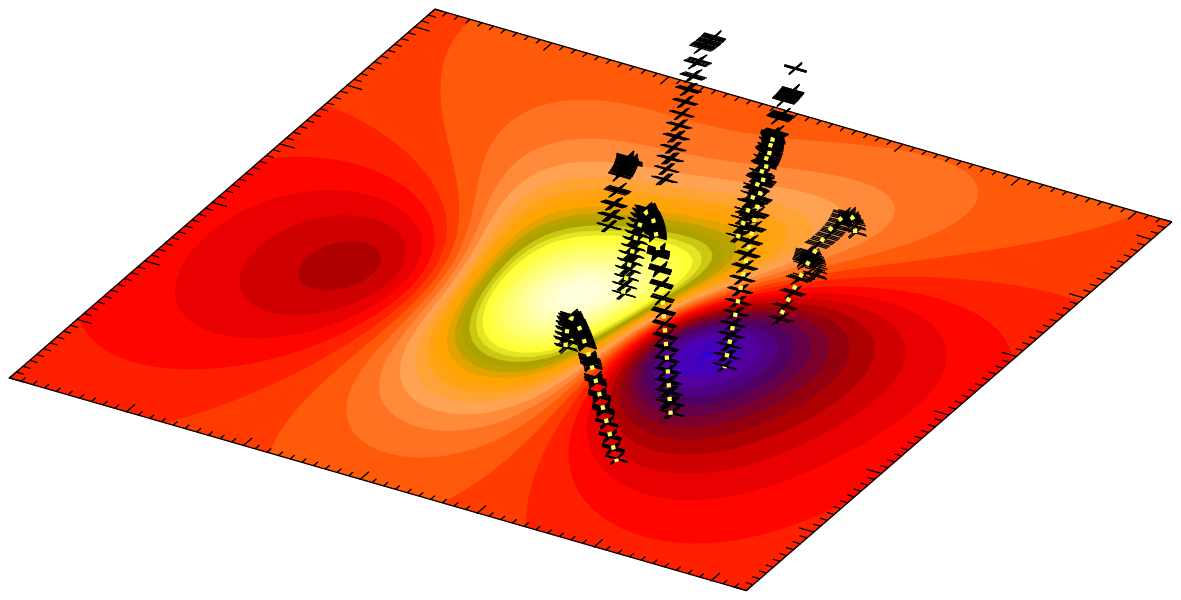}
\hspace{5mm}
\includegraphics[bb=190 50 300 190,clip,height=6cm,width=6cm]{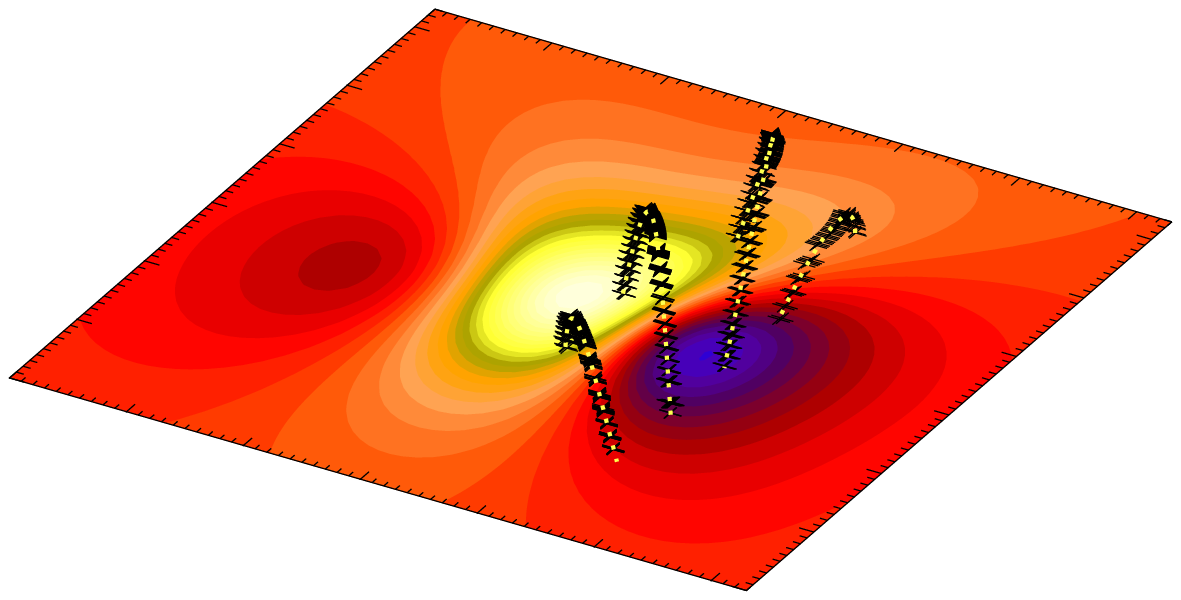}
\hspace*{\fill}\\
\caption{ Stereo image pair with identified loop projections (top). The
  coloured curves correspond to the projections of close field lines
  calculated from different magnetic field models which were used to
  establish the correspondences between the loop projections.
  A three-dimensional reconstruction of the observed loops includes
  ghosts if these correspondences are ignored (bottom left). The
  original model could faithfully be retrieved if they were accounted
  for (bottom right). From \citet{WiegelmannInhester:2006}.
  \label{fg:WgmInhfig5}}
\end{figure}

More sophisticated field assumptions postulate the vanishing of the
Lorentz force which is reasonable for quasistationary field
configurations given the dominance of the magnetic field in the
corona. These so-called ``force-free'' field models, however, require
the solution of a non-linear boundary value problem. The development
of codes which are able to effectively calculate force-free fields
from observed boundary is still in progress \citep{AmariEtal:2006,
InhesterWiegelmann:2006, SchrijverEtal:2006}.
A deficit of these models is that they require boundary values to
some extent not only on the solar surface where the observations
provides them but also on the computationally unavoidable lateral and
top boundaries where they have to be guessed. Hence, also the magnetic
field models available will not be unique and an intimate combination
of stereoscopic reconstruction and magnetic field modelling is probably
required to make definite progress.

With this approach in mind, \citet{GaryAlexander:1999} started from a
Laplace field extrapolation and suggested to gradually alter this initial
magnetic field until the projected field lines match best the observed
loop projections. Gary and Alexander considered a mere radial stretching
as deformation. The resulting
field is no more potential, it is not force-free either because the
radial stretching complies with toroidal currents which are not
necessarily aligned along the magnetic field.
In absence of stereo images, a magnetic field model could be produced
which agreed with observed loop projections in one image. The
resulting Lorentz force, however, necessary to achieve the agreement
by radial field line stretching was considerable.

\citet{LeeJKEtal:2006} proposed to make use of magnetic field information at a
very early stage of the image processing when loop projections are traced
in the images. As mentioned above, this segmentation of EUV images into
curves which trace out the loop projections is an image processing
problem of its own, but it is also a prerequisite for the tie-point
stereo reconstruction.
After a first step of identifying image pixels which lie on the
intensity ridge of a projected loop the magnetic field is used to
determine the connectivity of these ridge pixels.
In their oriented connectivity method,
Lee et al. restrict the range of azimuth orientations within which the
next ridge pixel is looked for from the range of magnetic field
azimuths that project onto the pixel from the Laplace field calculated
at various heights above the solar surface.

According to an idea of \citet{WiegelmannInhester:2006}, magnetic
field information could be used also to find correct correspondences
between loop projections in the stereo images. In a first step,
distances between projected loops in the stereo images and field line
projections onto these images are calculated. After appropriate
ordering of loops and field lines according to this distance those
loop projections in the two stereo images are assumed to correspond
which are closest to a common field line. The uniqueness constraint was
explicitly taken account of in this procedure and different distance
norms and field models were tested. However, this idea has so far only
been applied to simple synthetic loop models.
Fig.~\ref{fg:WgmInhfig5} shows at the top the projected
loops which were simulated from a known force-free magnetic field. The
closest field lines are superimposed with a colour code which relates to
different field line models used to find the corresponding loops.
Even the Laplace field model which can only insufficiently reproduce
the correct field still resulted in correct correspondences.

The magnetic field information could also be of help when the
reconstructed curve branches are connected across epipolar turning
points (see section~\ref{ch:ambiguities}). An estimate of the local
magnetic field direction in the vicinity of the turning point serve
a measure which connection is the most probable in a similar way
as in \citet{LeeJKEtal:2006}'s oriented connectivity method.

\section{Stereoscopy by tomography}\label{ch:tomography}

\begin{figure}
\hspace*{\fill}
\includegraphics[width=6cm]{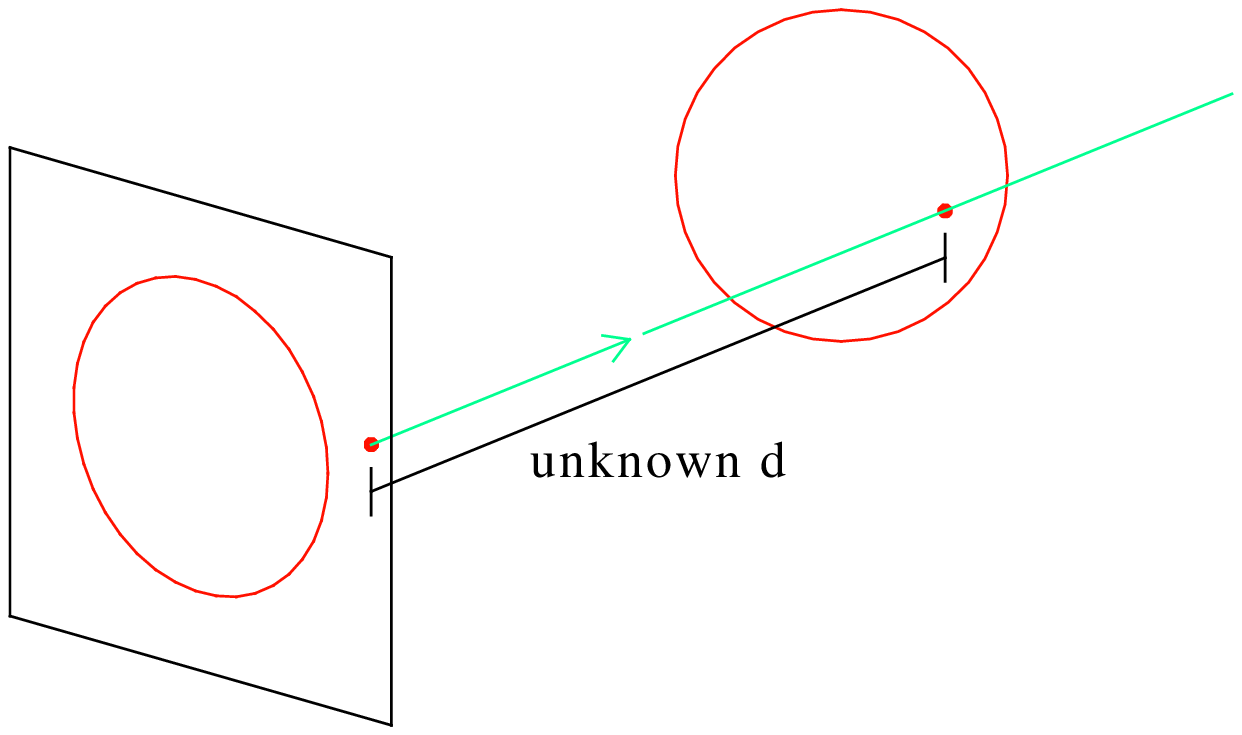}
\includegraphics[width=6cm]{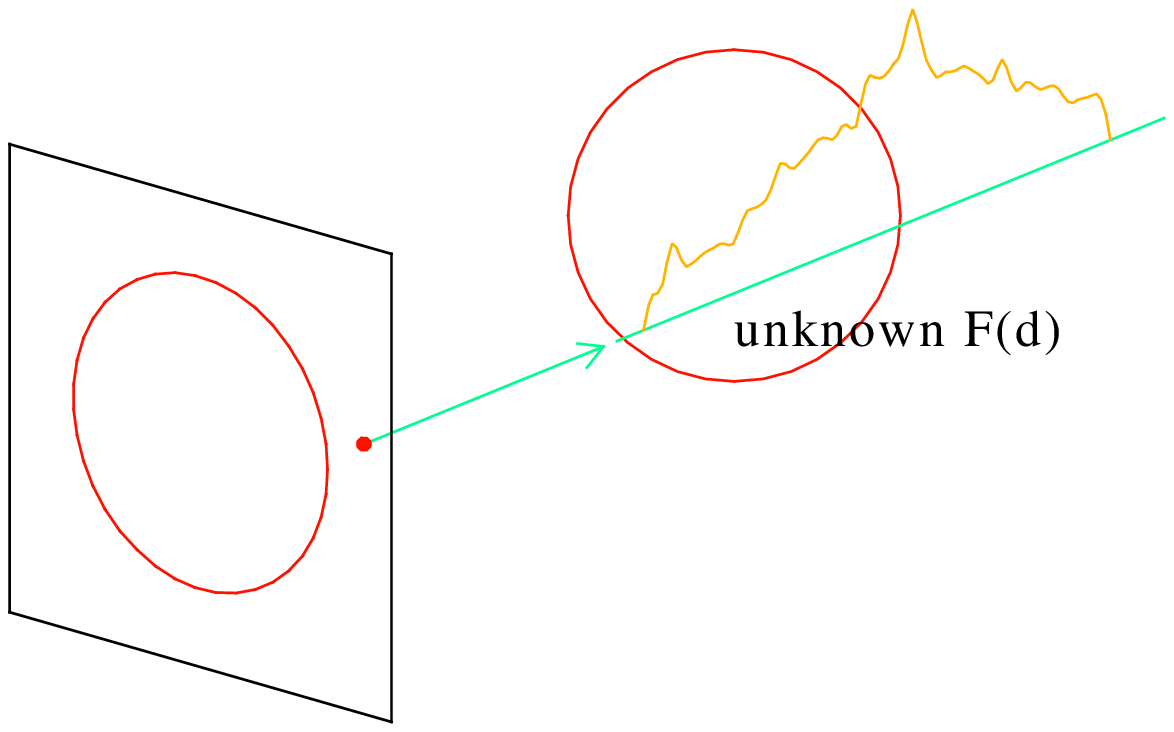}
\hspace*{\fill}
\caption{Assumed distribution of light sources along the projection ray
  from a bright image pixel in stereoscopy (left) and tomography (right).
  \label{fg:stereotomo}}
\end{figure}

Tomography provides a different approach to three dimensional
reconstruction. It differs from stereoscopy by a fundamental
assumption: Stereoscopy assumes that the light which illuminates an
individual image pixel is concentrated in a single space point while
in tomography the light source is assumed to be continuously
distributed along the line-of-sight inside a limited domain (see
Fig.~\ref{fg:stereotomo}).

Hence with stereoscopy, a second view is sufficient to locate the
isolated light source while a large number of additional views are
required to determine the continuous emissivity distribution.
However, if the source satisfies (more or less) the stereoscopy assumption
of locally concentrated light sources, a tomography scheme can still be used
because its assumptions are more general.

\begin{figure}
\hspace*{\fill}
\includegraphics[bb=117 43 508 799, clip, height=\textwidth, angle=-90]
                {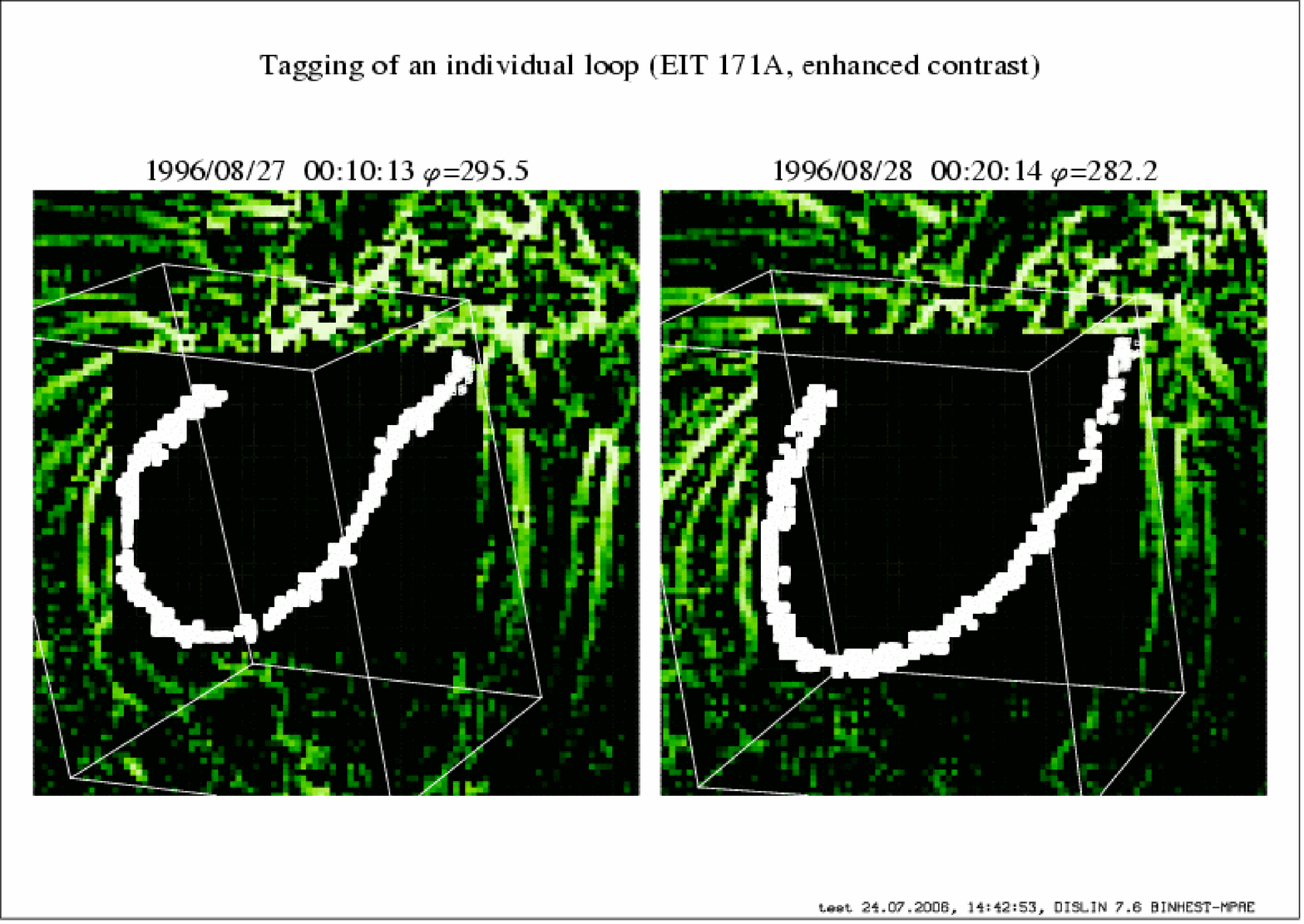}
\hspace*{\fill}
\caption{EIT observations of an active region taken 23 hours apart
  after contrast enhancement. In white we show the attempts to trace
  out a loop which is well visible in the contrast enhanced image.
\label{fg:stercon9a}}
\vspace*{10mm}
\hspace*{\fill}
\includegraphics[bb=43 49 510 517, clip, width=5cm]{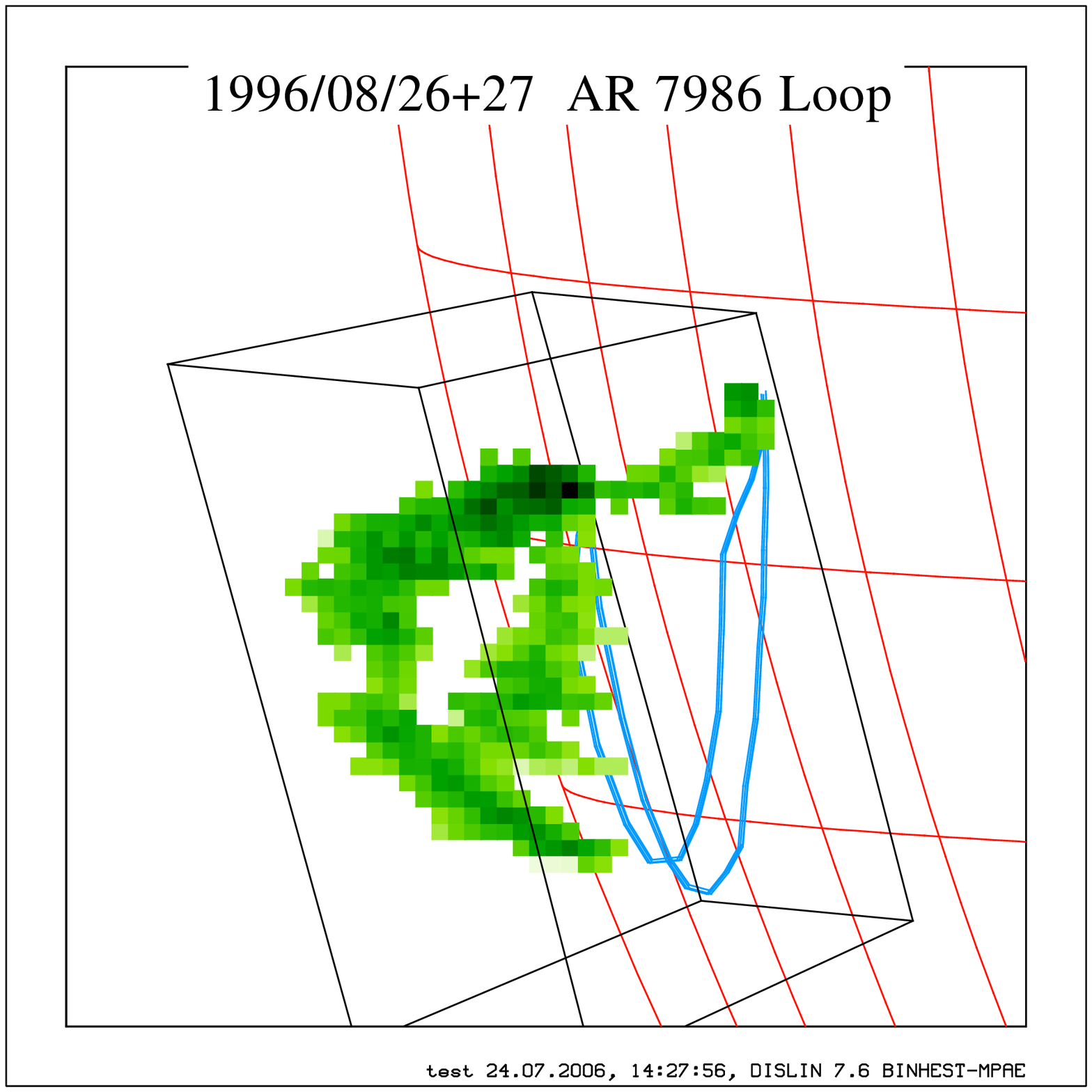}
\includegraphics[bb=43 49 510 517, clip, width=5cm]{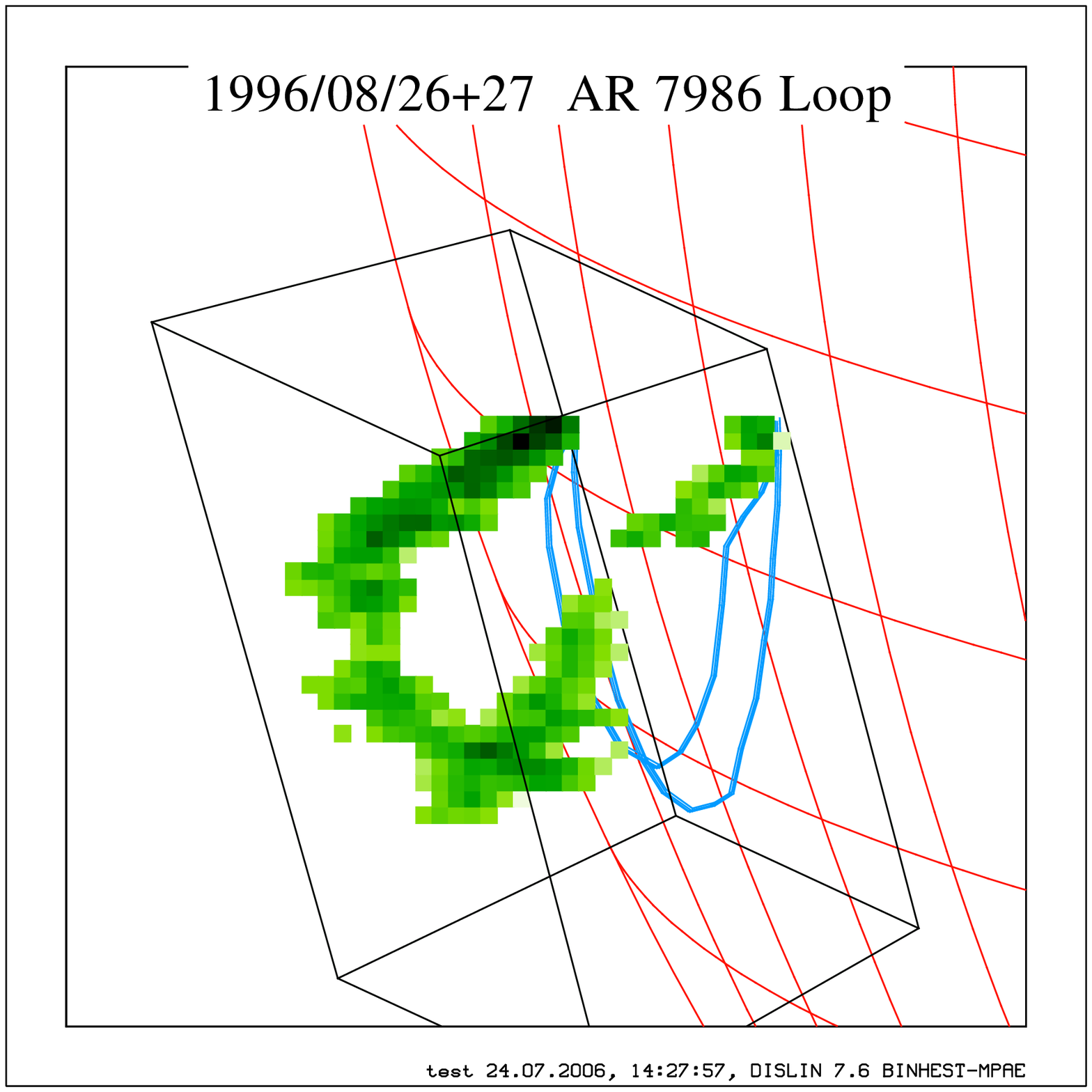}
\hspace*{\fill}\\
\hspace*{\fill}
\includegraphics[bb=43 49 510 517, clip, width=5cm]{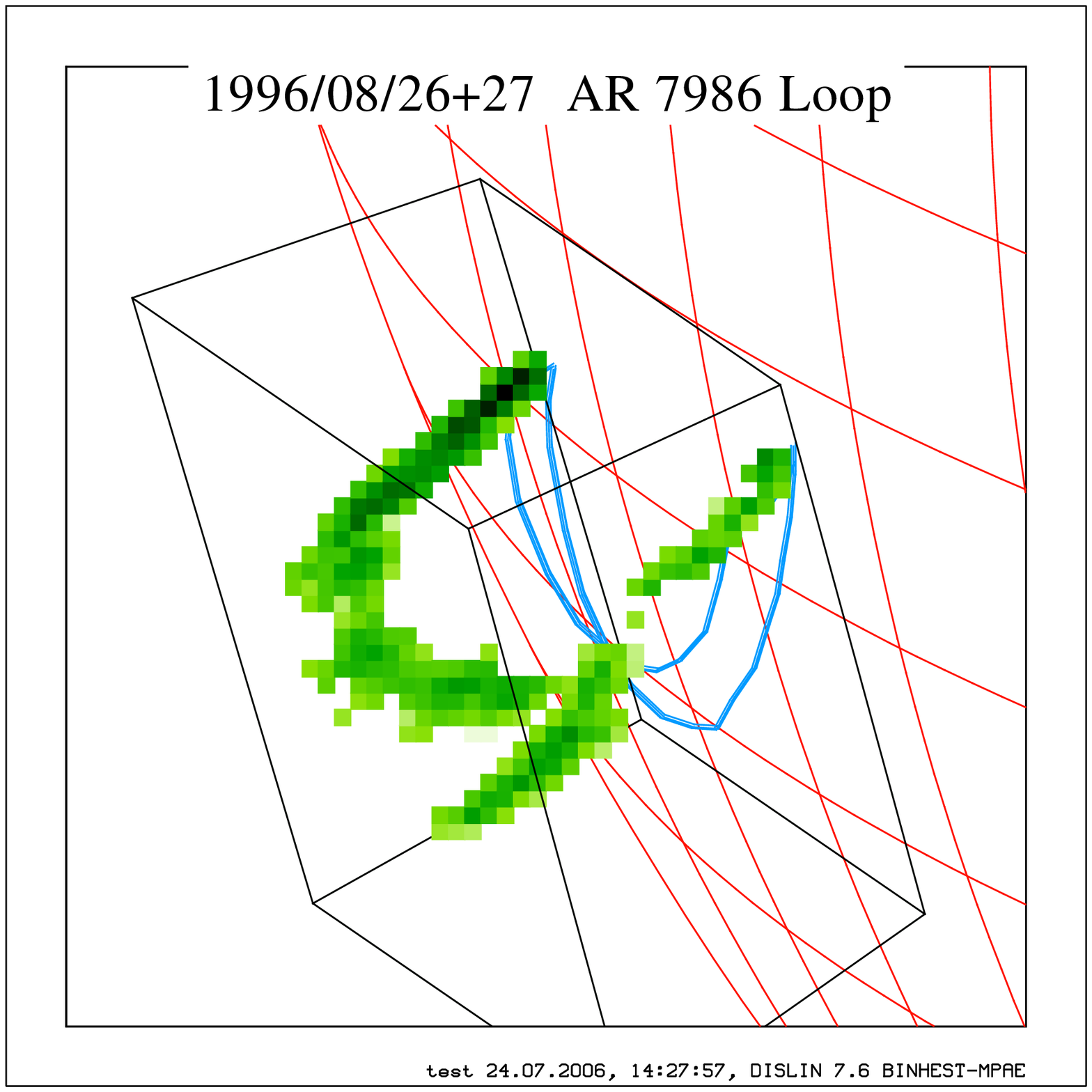}
\includegraphics[bb=43 49 510 517, clip, width=5cm]{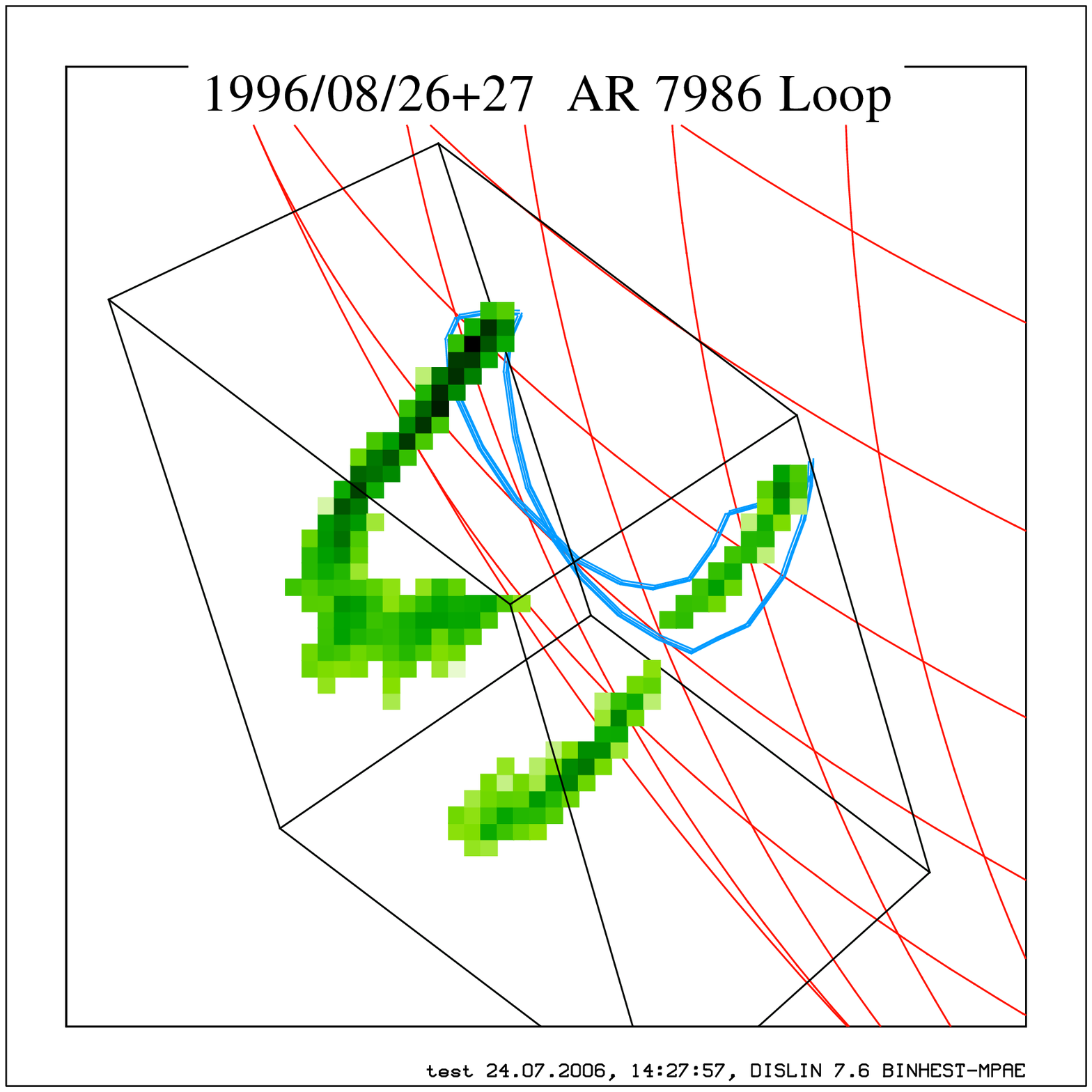}
\hspace*{\fill}
\caption{Tomographic reconstruction of a loop from the contrast enhanced
  data in Fig.~\ref{fg:stercon9a}.
\label{fg:stercon9b}}
\end{figure}

In tomography not only the location but also the intensity of loop
projections count and we can in principle hope to obtain a quantitative
emissivity distribution along the loop. The handicap of tomography
is, however, that it needs a larger set of view directions to resolve
the emissivity distribution. For the solar corona this can only be
obtained by exploiting the Sun's rotation and assuming stationarity of
the observed structures.

In Fig.~\ref{fg:stercon9a} and \ref{fg:stercon9b} we show an example
where we have tried to compromize between the additional requirements
and the additional output of tomography with respect to stereoscopy.
The intensity information is abandoned, the images are strongly
contrast enhanced and the background is thresholded to a nearly
black/white image.

The two stereo images were subsequently used as input for a tomography
inversion code. The zero background intensity after thresholding makes
this data ideal for a multiplicative inversion algorithms. We have used
a Richardson-Lucy algorithm (RLA, see, e.g., \citet{Lanteri+etal:1999}, for
alternatives see, e.g., \citet{Byrne:1998a}) to produce the results shown
in Fig.~\ref{fg:stercon9b}.

The original data is the same as for the tie-point reconstruction shown
in Fig.~\ref{fg:eit2}. Similar to this above example, our results here
suffer from the same deficiency: the small stereo base angle yields a large
uncertainty of the loop position in vertical direction which in
Fig.~\ref{fg:stercon9b} is visible by the relatively large spread of the
loop model in vertical direction.

It is also obvious that the tomography code just as the stereo
reconstruction is unable to distinguish between true and ghost
branches and some post-processing of the tomography results will in
general be necessary.
An advantage of the multiplicative algorithm we have used above is
that it converges after few iterations. If ghost branches can then be
identified in the resulting preliminary model distribution they just
need to be wiped out by setting the respective voxels to zero.
Resuming the iterations subsequently will redistribute the
uncompensated image intensities onto the remaining branches and the
eliminated ghosts will not reappear.
Another advantage of the tomography approach is that it can without
conceptual extensions cope with more than two images. All the geometrical
constraints which we must explicitly take care of in stereoscopy are
hidden in the sparse forward matrix which needs to be inverted for
tomography. This matrix grows, however, with every additional image.

\section*{Tie-point code example}

Stereoscopic reconstruction is in a way an inverse problem.
Working on problems of this kind requires some theoretical background
part of which we have tried to convey with this paper. A more concise
account of the geometrical  principles of stereoscopy can be found
in \citep{FaugerasLuong:2001}.
On the other hand, own experience is also indispensable when applying
your knowledge to a specific inverse problem. A review on modelling and
reconstruction attempts with solar physics data has been recently compiled
by \citep{ASchwandenEtal:2006}.
To enable the reader to make an own first step also in this direction,
we supply a simple tie-point code along with this paper.
It can be ordered as free software from the author
(email: {\tt binhest\@mps.mpg.de}).
You will recieve a file {\tt stereo\_02.tar.gz} which has to be unpacked in
the usual way. Installation instructions are given in the contained
{\tt readme} file. You further need a FORTRAN90/95 compiler and, if
you want to see the graphics output, the free DISLIN graphics library
from {\tt www.dislin.de}.

If you {\tt make} it, you produce an executable named {\tt stereo}.
This program reads observer parameters and 3D curve data from the file
{\tt stereo.dat}. The image projection is made in subroutine {\tt xc2pxc},
the image rectification in {\tt pxc2epi} and the reconstruction in {\tt
epi2xc}. The program can be run without graphics if you do not
have DISLIN. In this case you need to comment the graphics calls in
the source text in {\tt stereo.f90} accordingly and re{\tt make}.

If you use the graphics, the original curve {\tt xc} and its projections 
are drawn in red, the reconstructed curve branches {\tt xb} in yellow
(the original {\tt xc} should no more be visible). A left mouse click
relative to the cross in the image centre defines the direction into
which the main observer is to move for the subsequent image. This way
you can look at the reconstruction and its ghosts quasi interactively 
from all sides.
If you click on the \fbox{noise} box, the noise is added to the
projected curves on the images. The reconstruction is then redone from
the noisy curves. You may repeat this click to add more noise.
If you have enough of it, click onto the \fbox{stop}
box in the upper right.

You may produce your own curve in {\tt stereo.dat} and modify the
observer parameters. But you need to keep the obvious file layout in
order to enable {\tt stereo} to understand the data file. The
tie-point subroutines may of course also be extracted and incorporated into
your own analysis programs. Note, however that the author cannot take
any responsibility.

Please send questions, report bugs, suggestions and own improvements to
the author.

\section*{Acknowledgements}
The author thanks Thierry Dudoc de Witt for initiating and organizing
the ISSI working group meetings. Much of the work presented here grew
out of the discussions with him and the other working group members.
Thanks also for the hospitality of the ISSI during the working
meetings. I am also indebted to Li Feng, Borut Podlipnik, Ruan Peng,
Thomas Wiegelmann for helpful comments on an earlier version of the
manuscript.
The work was supported by DLR grant 50 OC 0501.

\bibliographystyle{issi}
\bibliography{BIStereoBasics}

\end{document}